\documentstyle[epsfig,twocolumn,subfigure]{mn}

\newcommand{\goodgap}{
\hspace{\subfigtopskip} \hspace{\subfigbottomskip}}

\title[Rotation Curves and $R^n$ gravity]{Analysis of Rotation Curves in the framework of $R^{n}$ gravity}
\author[C. Frigerio Martins and P. Salucci]{C. Frigerio Martins, P. Salucci\\
Astroparticle, Astrophysics Sectors, SISSA - Scuola Internazionale Superiore di Studi Avanzati,\\
Via Beirut, 4. 34014 Trieste, Italy\\
martins,salucci@sissa.it}

\date{Accepted xxx, Received yyy, in original form zzz}

\begin{document}

\maketitle

\begin{abstract}

We present an analysis of a devised sample of Rotation Curves (RCs), aimed at checking the consequences of a modified f(R) gravity on galactic scales.
Originally motivated by the the dark energy mystery, this theory may serve as a possibility of explaining the observed non-Keplerian profiles of galactic RCs in terms of a break-down of the Einstein General Relativity.
We show that in general the power-law f(R) version could fit well the observations with reasonable values for the mass model parameters, encouraging further investigation on $R^n$ gravity from both observational and theoretical points of view.

\end{abstract}


\section{Introduction}

It is well-known that the RCs of spiral galaxies show a non-Keplerian circular velocity profile which cannot be explained by considering a Newtonian gravitational potential generated by the baryonic matter \cite{persic}.
Current possible explanation of this controversy includes, among others, the postulate of a new yet not detected state of matter, the dark matter \cite{rubin83}, a phenomenological modification of the Newtonian dynamics \cite{milgrom,brownstein,sanders,bekenstein}, and higher order gravitational theories (originally devoted to solve the dark energy issue, see e.g., \cite{carroll,capozziello04}).

The recent theory proposed by Capozziello, Cardone $\&$ Troisi 2007 (hereafter CCT), modifies the usual Newtonian potential generated by baryonic matter in such a way that the predicted galaxy kinematics and the observed one have a much better agreement.
They consider power-law fourth order\footnote{The term comes from the fact that the generalized Einstein equations contain fourth order derivatives of the metric.} theories of gravity obtained by replacing in the gravity action the Ricci scalar $R$ with a function $f(R)\propto R^n$, where $n$ is a slope parameter. 
The idea is that the Newtonian potential generated by a point-like source gets modified in to
\begin{equation}
\phi(r) = -\frac{G m}{r} \{1+\frac{1}{2}[(r/r_c)^\beta-1]\} \label{eq: phi},
\end{equation}
where $\beta $ is a function of the slope $n$, and $r_c$ is a scale length parameter.
It turns out that in this theory $\beta$ is a universal constant while $r_c$ depends on the particular gravitating system being studied.
In a virialized system the circular velocity is related to the derivative of the potential through $V^2=r \:d\phi(r)/dr$.
It is clear that (\ref{eq: phi}) may help in the explanation of the circular velocity observed in spirals.

We remark that any proposed solution to the galaxy RC phenomenon must not only fit well the kinematics but, equally important, also have best-fit values of the mass model parameters that are consistent with well studied global properties of galaxies.

For a sample of 15 Low Surface Brightness galaxies the model described in CCT  was fairly able to fit the RCs. 
However, in our view, the relevance of their finding is limited by the following considerations: 
\begin{itemize}
\item the sample contains several objects whose RCs are not smooth, symmetric and extended to large radii
\item the sample contains only Low Surface Brightness galaxies while a wider sample is desirable 
\item the universal parameter $n$ is not estimated by the analysis itself but it is taken from other observations. 
\end{itemize}  

In the present work we generalize the results of Capozziello \emph{et al.} 2007 and test a wider and fairer sample of spirals, improving the analysis methodology.
Our goal is to perform a check of their model on galactic scales in order to investigate its consistency and universality.

The plan of this article is the following: in Sect.2 we briefly summarize the main theoretical results described in CCT relevant for the analysis of our sample.
In Sect.3 we present our sample and methodology of analysis.
In Sect.4 the results are presented and finally the conclusions in Sect.5. 

\section{Newtonian limit of $f(R)$ gravity}

The theory proposed by Capozziello \emph{et al.} 2007 is an example of $f(R)$ theory of gravity \cite{nojiri,carloni}.
In these theories the gravitational action is defined to be:
\begin{equation}
{\cal S}=\int d^4 x \: \sqrt{-g} \:[f(R)+{\cal L}_m] \label{eq: action}
\end{equation}
where $g$ is the metric determinant, $R$ is the Ricci scalar  and ${\cal L}_m$ is the matter Lagrangian.
They consider:
\begin{equation}
f(R)=f_0 R^n \label{eq: f}
\end{equation}
where $f_0$ is a constant to give correct dimensions to the action and $n$ is the slope parameter.
The modified Einstein equation is obtained by varying the action with respect to the metric components.

Solving the vacuum field equations for a Schwarzschild-like metric in the Newtonian limit of weak gravitational fields and low velocities, the modified gravitational potential for the case of a point-like source of mass $m$, is given by (\ref{eq: phi}), where the relation between the slope parameter $n$ and $\beta$ (see detailed calculation in CCT) is given by:
\begin{equation}
\beta = \frac{12 n^2 -7 n - 1 - \sqrt{36 n^4 + 12 n^3 - 83 n^2 + 50 n + 1}}{6 n^2 -4 n + 2}. \label{eq: beta}
\end{equation}
Note that for $n=1$ the usual Newtonian potential is recovered.  
The large and small scale behavior of the total potential constrain the parameter $\beta$ to be $0 < \beta < 1$.

The solution (\ref{eq: phi}) can be generalized to extended systems with a given density distribution $\rho(r)$ by simply writing:
\begin{eqnarray}
\phi(r) & = & -G \int d^{3}r' \:\frac{\rho(\textbf{r'})}{|\textbf{r}-\textbf{r'}|}\:\:\{1+\frac{1}{2}[\frac{|\textbf{r}-\textbf{r'}|^\beta}{r_{c} ^\beta}-1]\}\nonumber\\
~ & = &\phi_{N}(r)+\phi_{C}(r) \label{eq:tot},
\end{eqnarray}
where $\phi_{N}(r)$ represents the usual Newtonian potential and $\phi_{C}(r)$ the additional correction.
In this way, the Newtonian potential can be recuperated when $\beta=0.$
The solution for the specific density distribution relevant for spiral galaxies is described in the following paragraph.

\section{Data and Methodology of the test}
We selected two samples of galaxies: a first with 15 galaxies, called \emph{Sample A}, that represents the best available RCs to study the mass distribution of luminous and/or dark matter, and it has been used in works concerning modifications of gravity and the  core/cusp controversy \cite{corbelli,gentile,frigerio}.

This sample includes nearby galaxies of different Surface Brightness: DDO 47 \cite{47}; ESO 116-G12, ESO 287-G13, NGC 7339, NGC 1090 \cite{gentile}; UGC 8017, UGC 10981, UGC 11455 \cite{vogt}; M 31, M 33 \cite{corbelli}; IC 2574 \cite{2574}, NGC 5585 \cite{5585}, NGC 6503 \cite{6503}, NGC 2403 \cite{2403}, NGC 55 \cite{55}.
This sample is the most suitable for a fair test of theories like the one of Capozziello \emph{et al.} 2007:
\begin{itemize}
\item The RCs are smooth, symmetric and extended to large radii.
\item The galaxies present a very small bulge so that it can be neglected in the mass model to a good approximation.
\item The luminosity profile is well measured and presents a smooth behavior
\item The data are uniform in quality up to the maximal radii of each galaxy. 
\end{itemize}
Let us notice that in some of these galaxies H$_\alpha$ and HI RCs are  both available and in these cases they  agree well where they coexist.

We also considered a second sample called \emph{Sample B} consisting of 15 selected objects from Sanders \& McGaugh 2002 that has been used to test MOND.
This sample consists of the following galaxies: UGC 6399, UGC 6983, UGC 6917, NGC 3972, NGC 4085, NGC 4183, NGC 3917, NGC 3949, NGC 4217, NGC 3877, NGC 4157, NGC 3953, NGC 4100 \cite{umajorPhotometry,umajorHI}; NGC 300 \cite{300}; UGC 128 \cite{128}. 
Although these galaxies do not fulfill all the requirements of \emph{Sample A} we have analyzed them for completeness sake.
The properties of the galaxies of the two samples are listed in Table 1.
Notice that the theory of Capozziello \emph{et al.} 2007 requires an analysis with a sample of high quality galaxies, as described above, where each luminous profile plays an important role, whereas this is not the case in MOND. 

We decompose the total circular velocity into stellar and gaseous contributions.
Available photometry and radio observations show that the stars and the gas in our sample of galaxies are distributed in an infinitesimal thin and circular symmetric disk.
While the HI surface luminosity density distribution $\Sigma_{gas}(r)$ gives a direct measurement of the gas mass, optical observations show that the stars have an exponential distribution:
\begin{equation}
\Sigma_{D}(r)=(M_{D}/2 \pi R_{D}^{2})\: e^{-r/R_{D}}\label{eq:sigma},
\end{equation}
where $M_{D}$ is the disk mass and $R_{D}$ is the scale length, the latter being measured directly from the optical observations, while $M_D$ is kept as a free parameter of our analysis.

The distribution of the luminous matter in spiral galaxies has to a good extend cylindrical symmetry, hence using cylindrical coordinates, the potential (\ref{eq:tot}) reads
\begin{equation}
\phi(r)=-G\int^{\infty}_{0}dr'\:r'\Sigma(r')\int^{2\pi}_{0} \frac{d\theta}{|\textbf{r}-\textbf{r'}|}\{1+\frac{1}{2}[\underbrace{\frac{|\textbf{r}-\textbf{r'}|^\beta}{r_{c} ^\beta}}-1]\}.\label{eq:cyl}
\end{equation}
$\Sigma(r')$ is the surface density distribution of the stars, given by (\ref{eq:sigma}) , or of the gas, given by an interpolation of the HI data points up to the last measured point.
$\beta$ and $r_c$ are free parameters of the theory, with the latter galaxy dependent.
We neglected the gas contribution to the mass density for radii larger than the last measured point, however we checked the goodness of this approximation by extending the distribution with a different kind of decreasing smooth curves and realized that error made in the truncated approximation is small enough to be neglected.

Defining $k^{2}\equiv \frac{4r\:r^{'}}{(r+r^{'})^2}$, we can express the distance between two points  in cylindrical coordinates as $|\textbf{r}-\textbf{r'}|=(r+r)^2 (1-k^2cos^2(\theta/2))$.
The derivation of the circular velocity due to the marked term of equation (\ref{eq:cyl}), that we call $\phi_{\beta}(r)$, is now direct:
\begin{equation}
r\:\frac{d}{dr}\:\phi_{\beta}(r) =  -2^{\beta -3} r^{-\beta}_{c}\: \pi \: \alpha \: (\beta -1)\: G\:I(r) \label{eq: vel}, 
\end{equation}
where the integral is defined as
\begin{equation}
{\cal{I}}(r) \equiv \int^{\infty}_{0} dr' r' \frac{\beta -1}{2}k^{3-\beta}\: \Sigma(r')\:{\cal{F}}(r) \label{eq: int},
\end{equation}
with ${\cal{F}}(r)$ written in terms of confluent hyper-geometric function: ${\cal{F}}(r) \equiv 2(r+r')\:_{2} F_{1}[{\frac{1}{2},\frac{1-\beta}{2}},{1},k^{2}]+[(k^{2}-2)r'+k^2 r]\:_{2} F_{1}[{\frac{3}{2},\frac{3-\beta}{2}},{2},k^{2}]$.

The total circular velocity is the sum of each squared contribution:  
\begin{equation}
V_{CCT}^{2}(r)=\:V^{2}_{N,stars}+V^{2}_{N,gas}+V^{2}_{C,stars}+V^{2}_{C,gas}\label{eq: vtot}
\end{equation}
where the $stars$ and $gas$ subscripts refer to the different contributions of luminous matter to the total potential (\ref{eq:tot}). The \emph{N} and \emph{C} subscripts refer to the Newtonian and the additional correction potentials.

Let us recall that we can write
\begin{equation}
V_{N,stars}^{2}(r)=(G M_{D}/2R_{D}) \:  x^{2}B(x/2)\label{eq: vN},
\end{equation}
where $x\equiv r/R_{D}$, $G$ is the gravitational constant and the quantity $B=I_{0}K_{0}-I_{1}K_{1}$ is a combination of Bessel functions \cite{freeman}.

Galaxies UGC 8017, M 31, UGC 11455 and UGC 10981 presents a very small amount of gas and for this reason it has been neglected in the analysis.
Notice that the correction to the Newtonian potential in equation (\ref{eq: phi}) may be negative and this would lead to a negative value of $V^2_C$.
In Figures 1 and 2 however the velocities $V_C$ are shown only in the ranges of $r$  where their square are positive.

In a first step, the RCs are $\chi^{2}$ best-fitted with the following free parameters: the slope ($\beta$) and the scale length ($r_{c}$) of the theory, and the gas mass fraction ($f_{gas}$) related to the disk mass simply by $M_{D}=M_{gas}(1-f_{gas})/f_{gas}$.
The errors for the best fit values of the free parameters are calculated at one standard deviation with the $\chi^2_{red}+1$ rule.
From the results of these fits we get a mean value of $\beta=0.7 \pm 0.25$ ($n\simeq2.2$). 
In the second step we redo the best-fit fixing the slope parameter at $\beta =0.7$ keeping as free parameters only $r_c$ and $f_{gas}$.
Notice that in a previous paper \cite{capozziello06}, a mean value of  $\beta=0.58 \pm 0.15$ ($n\simeq1.7$)  has been obtained, perfectly compatible with our result.
This parameter however, is well constrained from SNeIa observations to be $\beta=0.87$ ($n\simeq3.5$), also compatible with our measurements. 
In our analysis the value $\beta=0.7$ is the most favorable for explaining the RCs: different values of $\beta$ from the one we adopt here lead to worse performance.

\section{Results}

We summarize the results of our analysis in Figures 1 and 2 and Table 1\footnote{Numerical codes and data used to obtain these results can be found at the address http://people.sissa.it/$\sim$martins/home.html}.
In general we find for all galaxies:
\begin{itemize}
\item the velocity model $V_{CCT}$ well fitting the RCs 
\item acceptable values for the stellar mass-to-light ratio
\item too vast range for values of the gas fraction (0$\% < f_g  < 100\%$)
\item not clear comprehension for the big variation of values for the scale length parameter (0.005 kpc$<r_c<$1.53 kpc).
\end{itemize}

The residuals of the measurements with respect to the best-fit mass model are in most of the cases compatible with the error-bars, see Figures 1 and 2, though three galaxies show significant deviations: NGC 6503, NGC 2403 and M 33.
We also find acceptable values for the B-band mass-to-light ratio parameter for most of the galaxies, for which we should have approximately $0.5<\Upsilon_{\star}^{B}<6$ and a positive correlation between B-luminosity\footnote{$\Upsilon_{\star}^B \equiv M_{D}/L_{B}$; $M_D$ is the disk mass and $L_{B}$ is the B-band galaxy luminosity} and $\Upsilon_{\star}^{B}$ \cite{salucci}:
\begin{eqnarray}
M_D(L_B) & \simeq & 3.7 \times 10^{10} \nonumber \\
&  \times & [(\frac{L_B}{L_{10}})^{1.23} \: g(L_B)+0.095(\frac{L_B}{L_{10}})^{0.98}]M_{\odot}\label{eq: salucci99},
\end{eqnarray}
where $L_{10}\equiv 10^{10} L_{B \odot}$ and $g(L_B)=exp[-0.87 \times (log \frac{L_B}{L_{10}}-0.64)^2]$.
In detail we find discrepancies for NGC 55, UGC 8017, NGC 3972, NGC 4085 and NGC 4183.
Values for the scale length parameter ($r_c$) are in general smaller for less massive galaxies and bigger for more massive ones.
We obtained a Newtonian fit for UGC 10981, as shown by the exceedingly large value for $r_c$, see Figure 1.

The model analyzed in this article yields better results on galactic scales than $\Lambda$ Cold Dark Matter models, where in the latter these galaxies have serious problems like marginal fits and unreasonable values for the stellar mass-to-light ratio, see e.g., Frigerio Martins \& Salucci 2007 and Gentile \emph{et al.} 2004.

\section{Conclusions}

We have investigated the possibility of fitting the RCs of spirals with a power-low fourth order theory of gravity of Capozziello \emph{et al.} 2007, without the need of dark matter.
We remark the relevance of our sample that contains objects in a large range of luminosity and with very accurate and proper kinematic.
We find in general a reasonable agreement, with some discrepancies, between the RCs and the Capozziello \emph{et al.} 2007 circular velocity model, encouraging further investigations from the theoretical point of view.

\textit{Acknowledgments}. We warmly thank S. Capozziello, J. Miller and T. Sotiriou for useful discussions. 
This research was supported by CAPES-Brasil (C.F.M).

\begin{figure*}
\centering
\subfigure{\includegraphics[width=4.75cm]{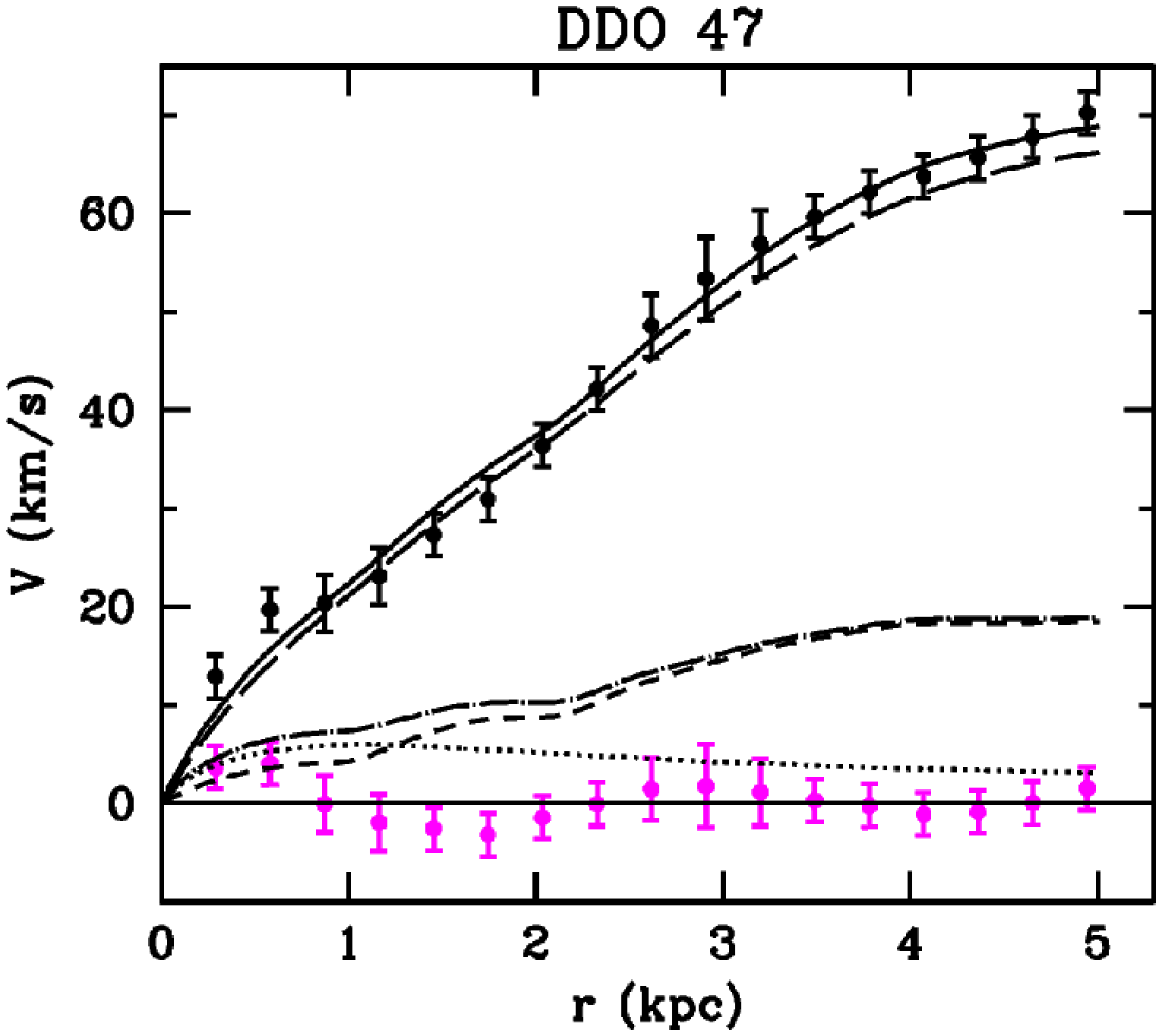}}\goodgap
\subfigure{\includegraphics[width=4.75cm]{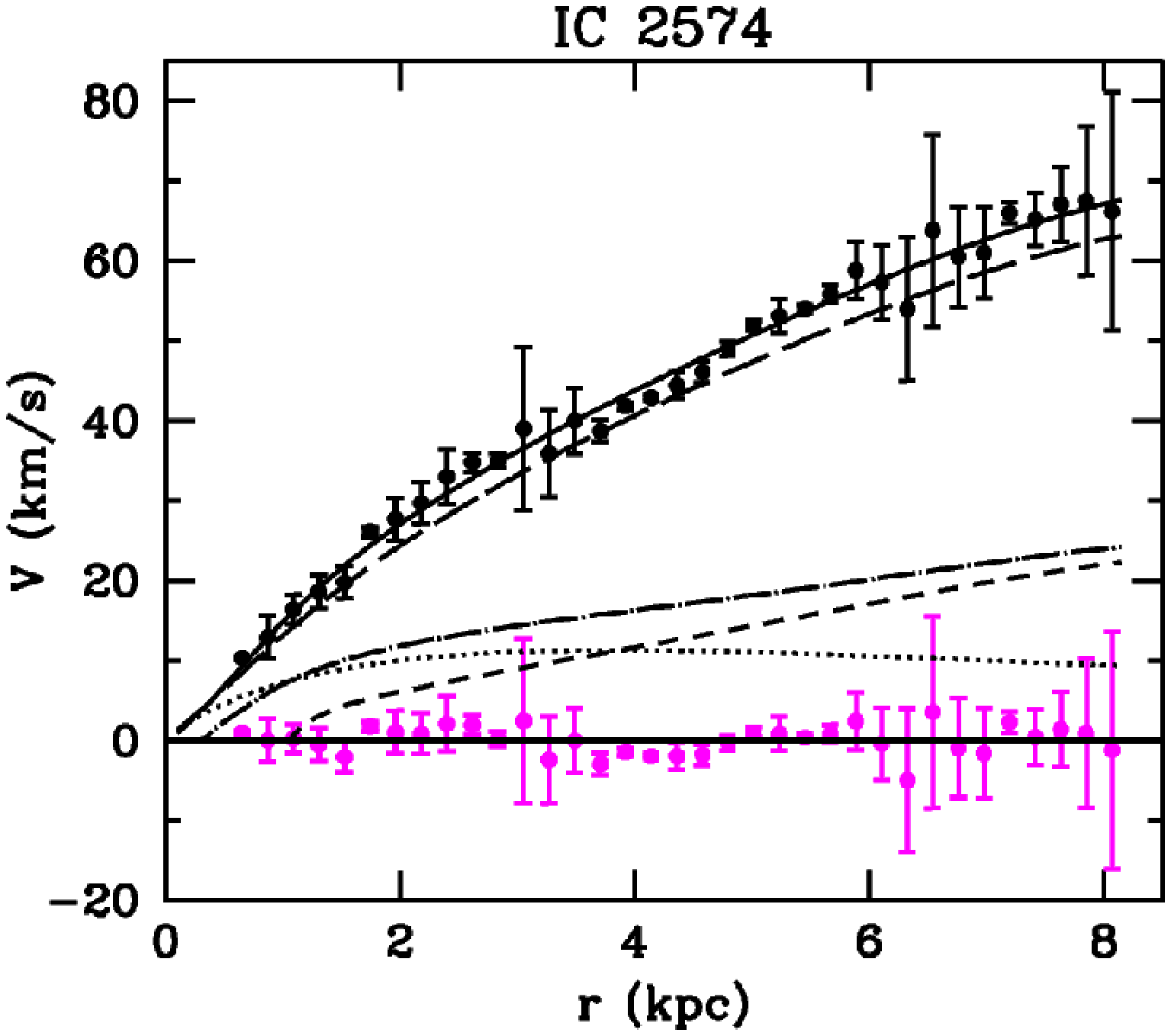}}\goodgap
\subfigure{\includegraphics[width=4.75cm]{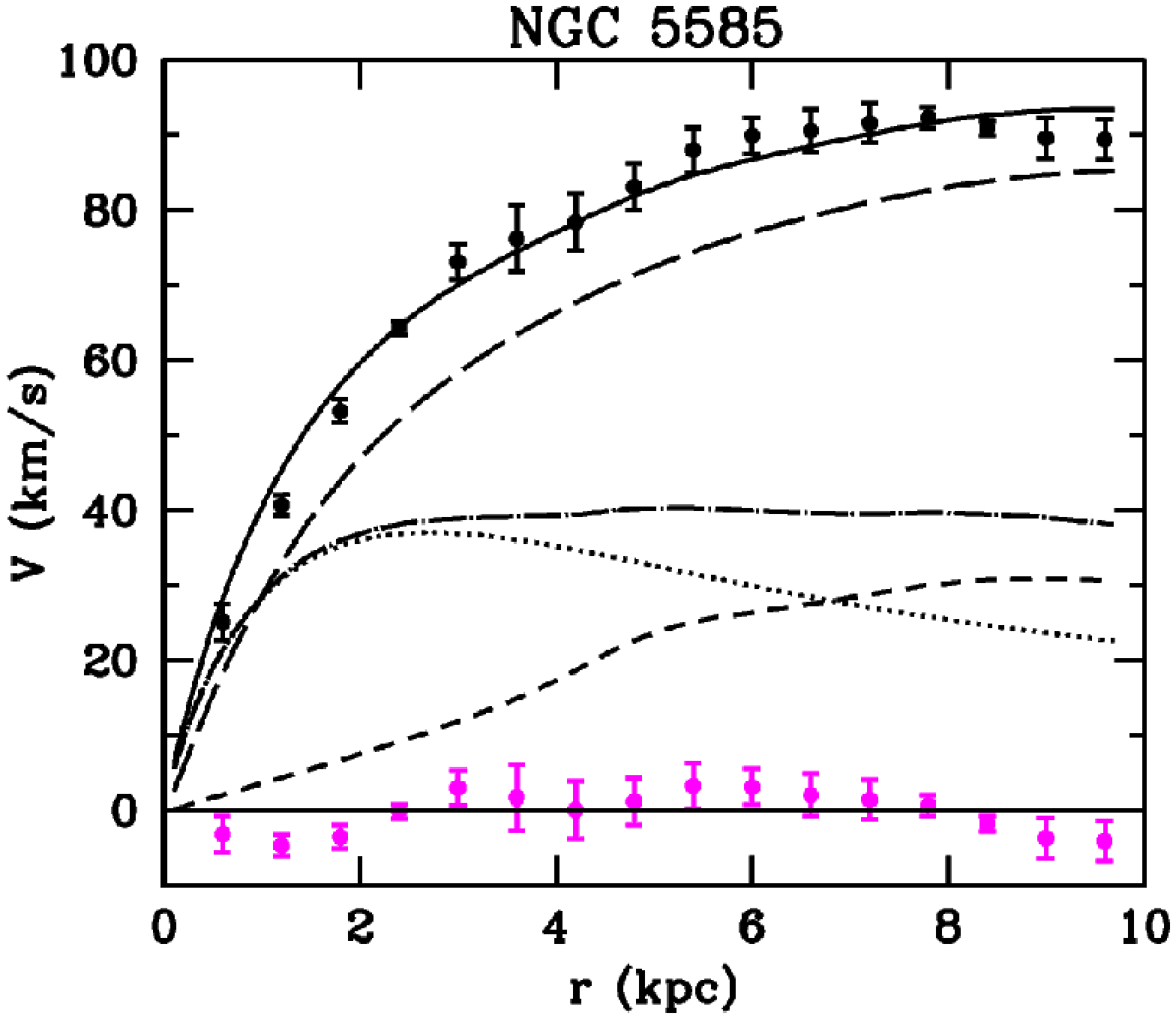}}\goodgap\\
\subfigure{\includegraphics[width=4.75cm]{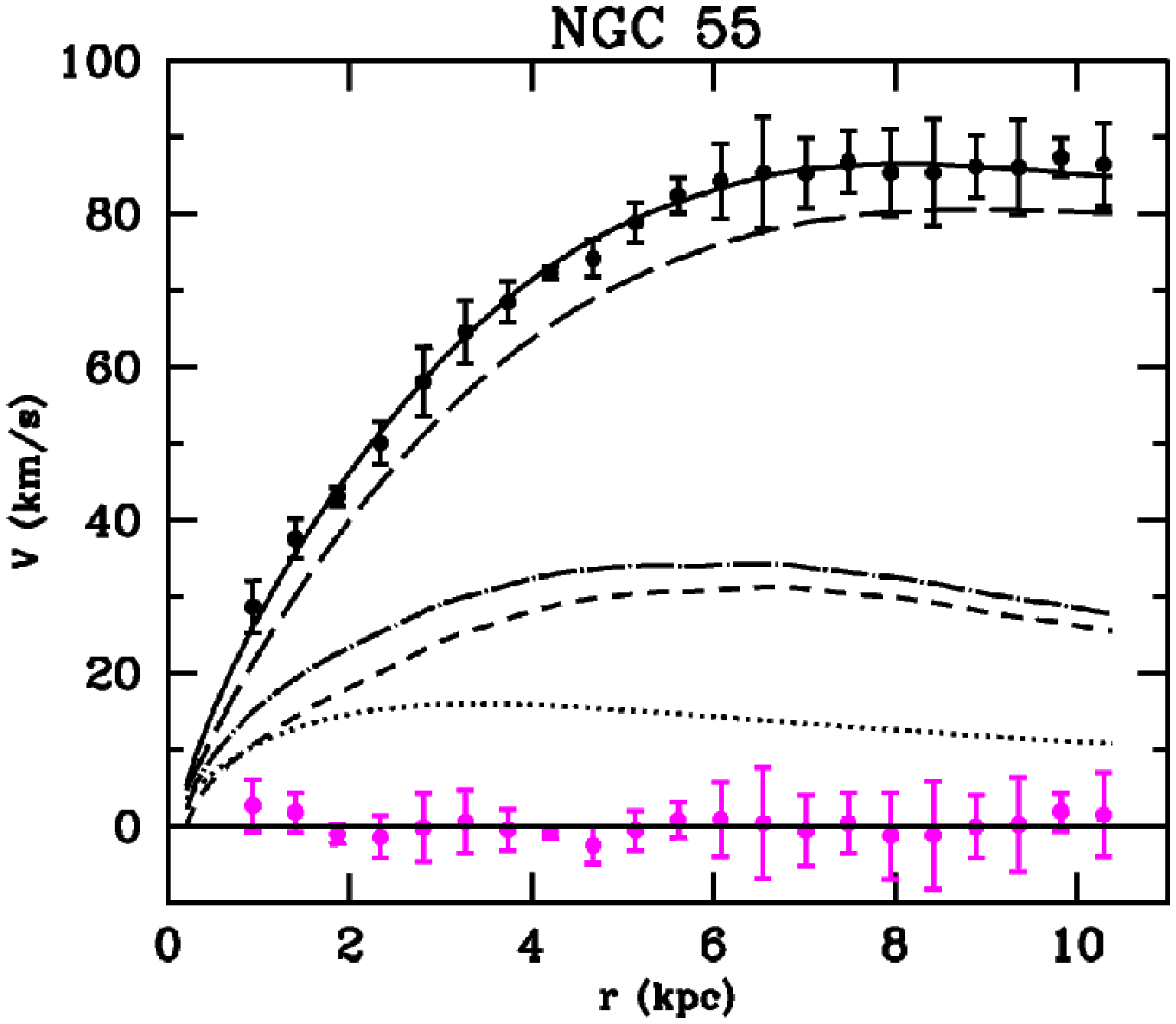}}\goodgap
\subfigure{\includegraphics[width=4.75cm]{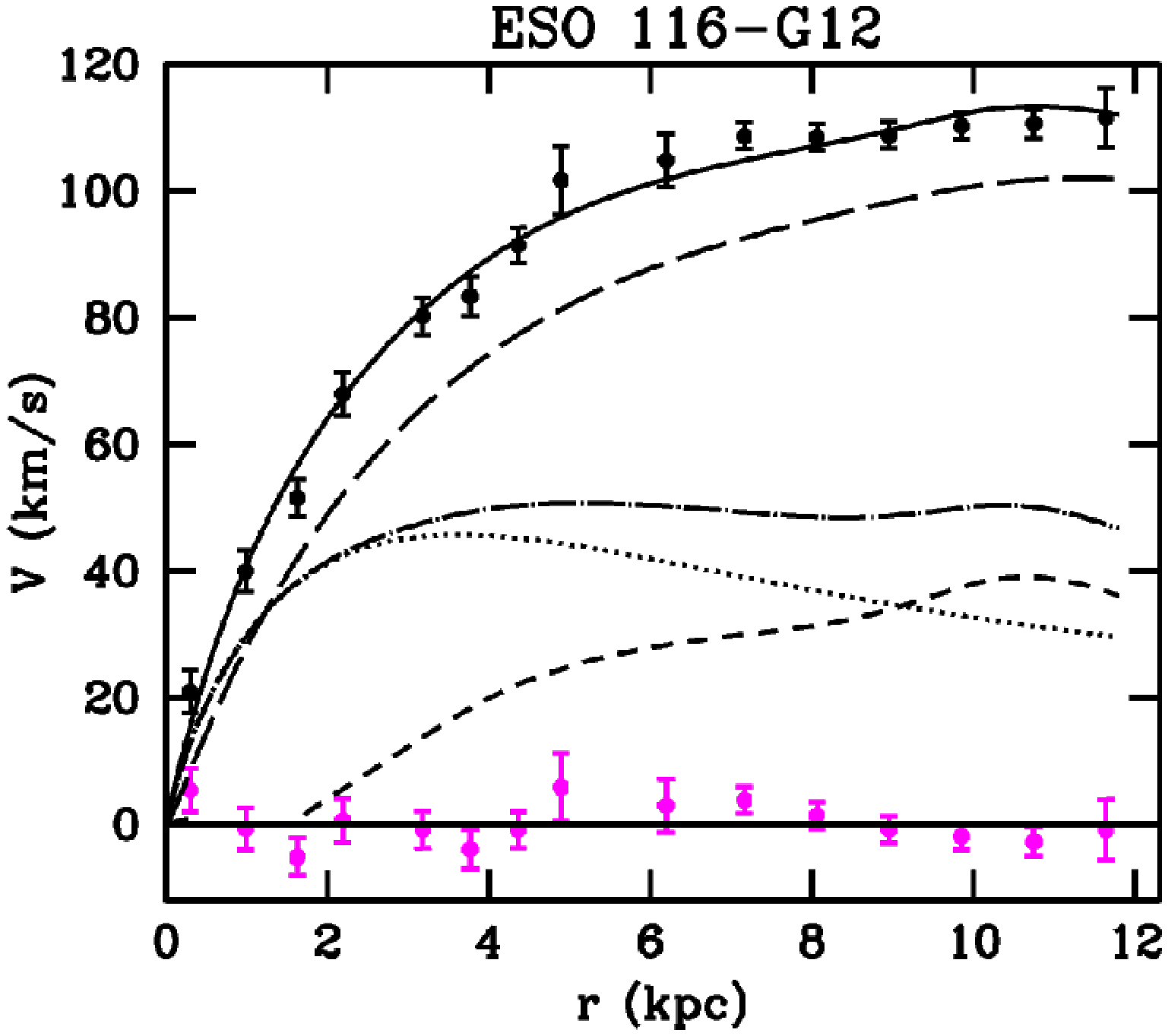}}\goodgap
\subfigure{\includegraphics[width=4.75cm]{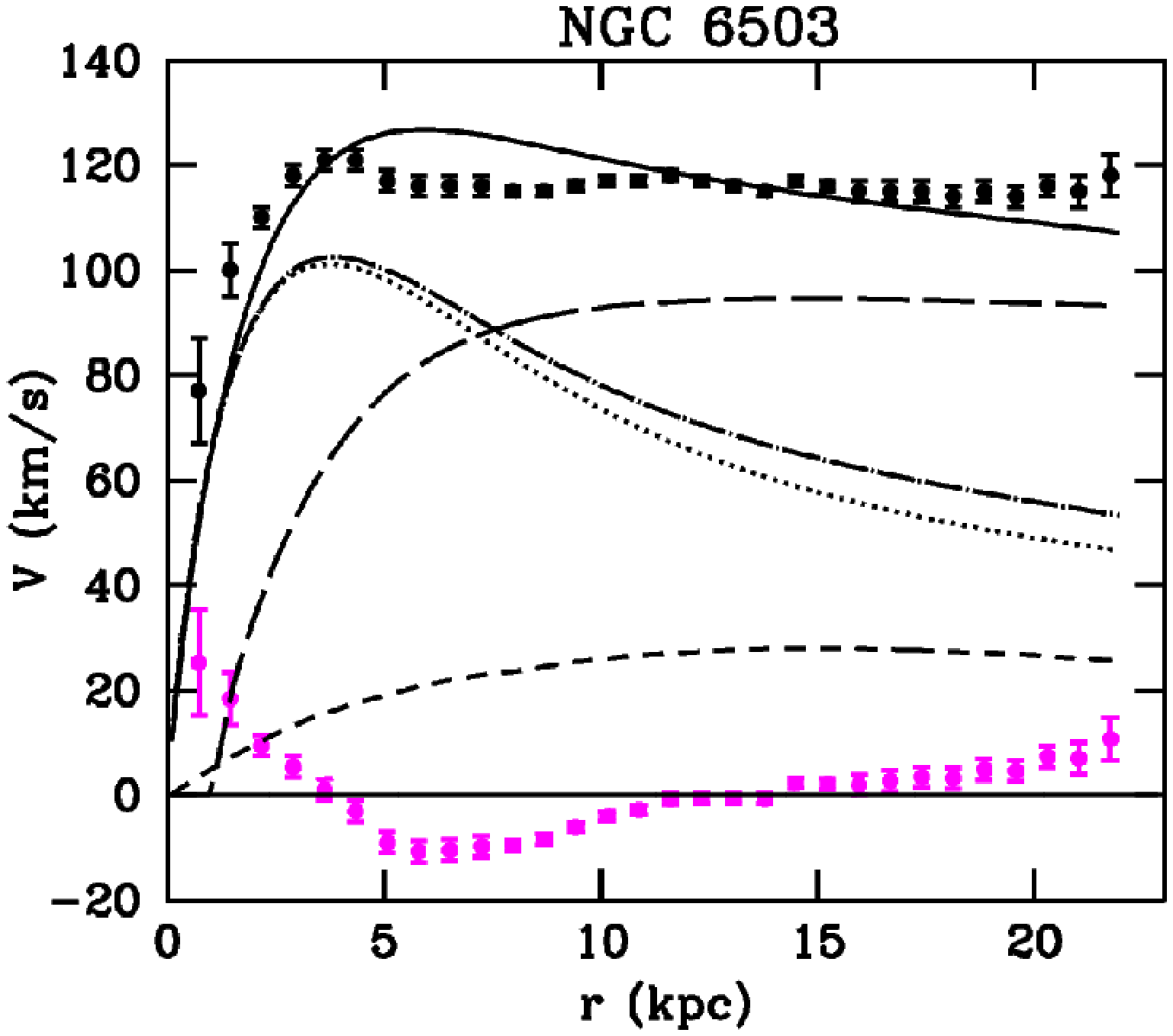}}\goodgap\\
\subfigure{\includegraphics[width=4.75cm]{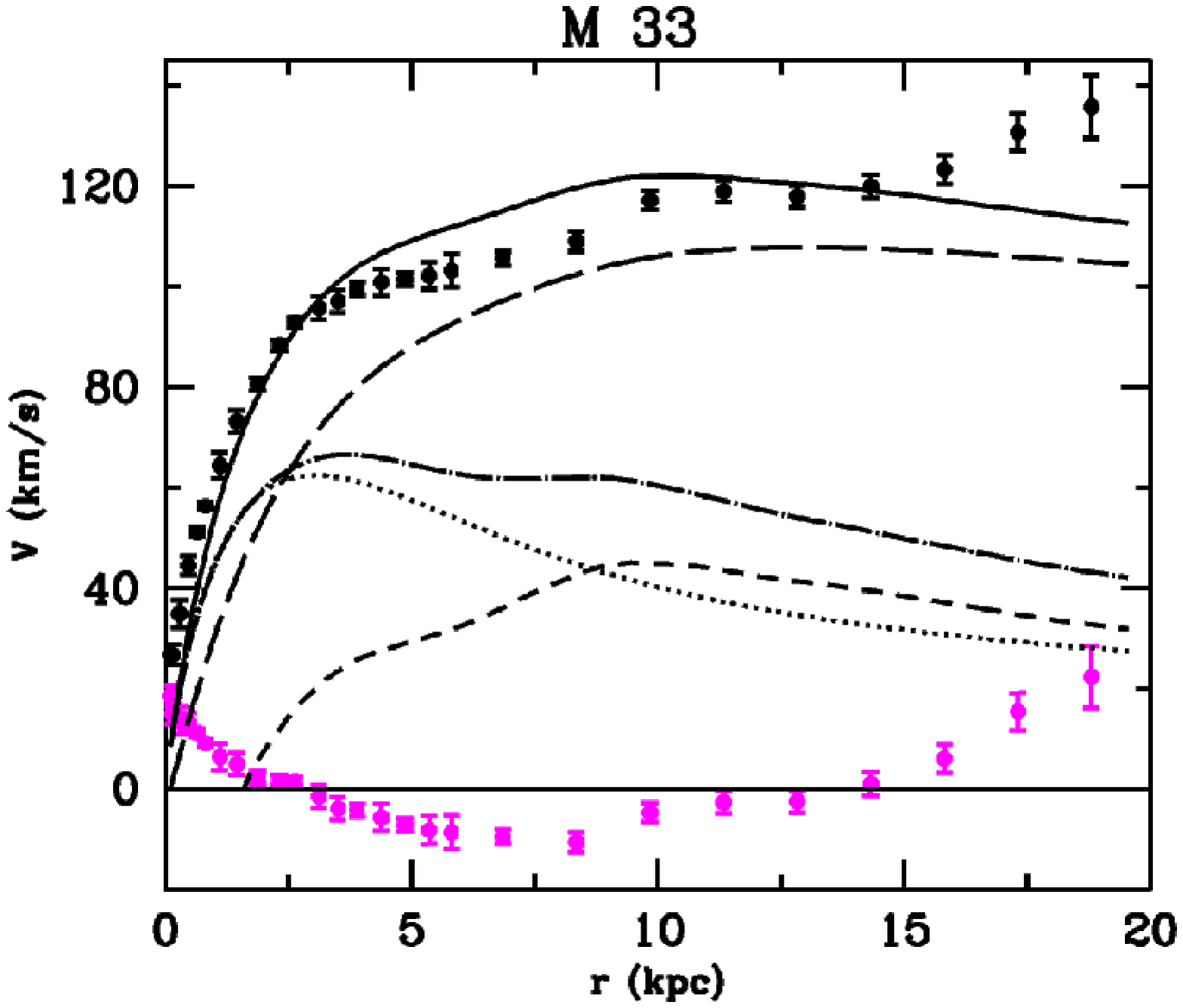}}\goodgap
\subfigure{\includegraphics[width=4.75cm]{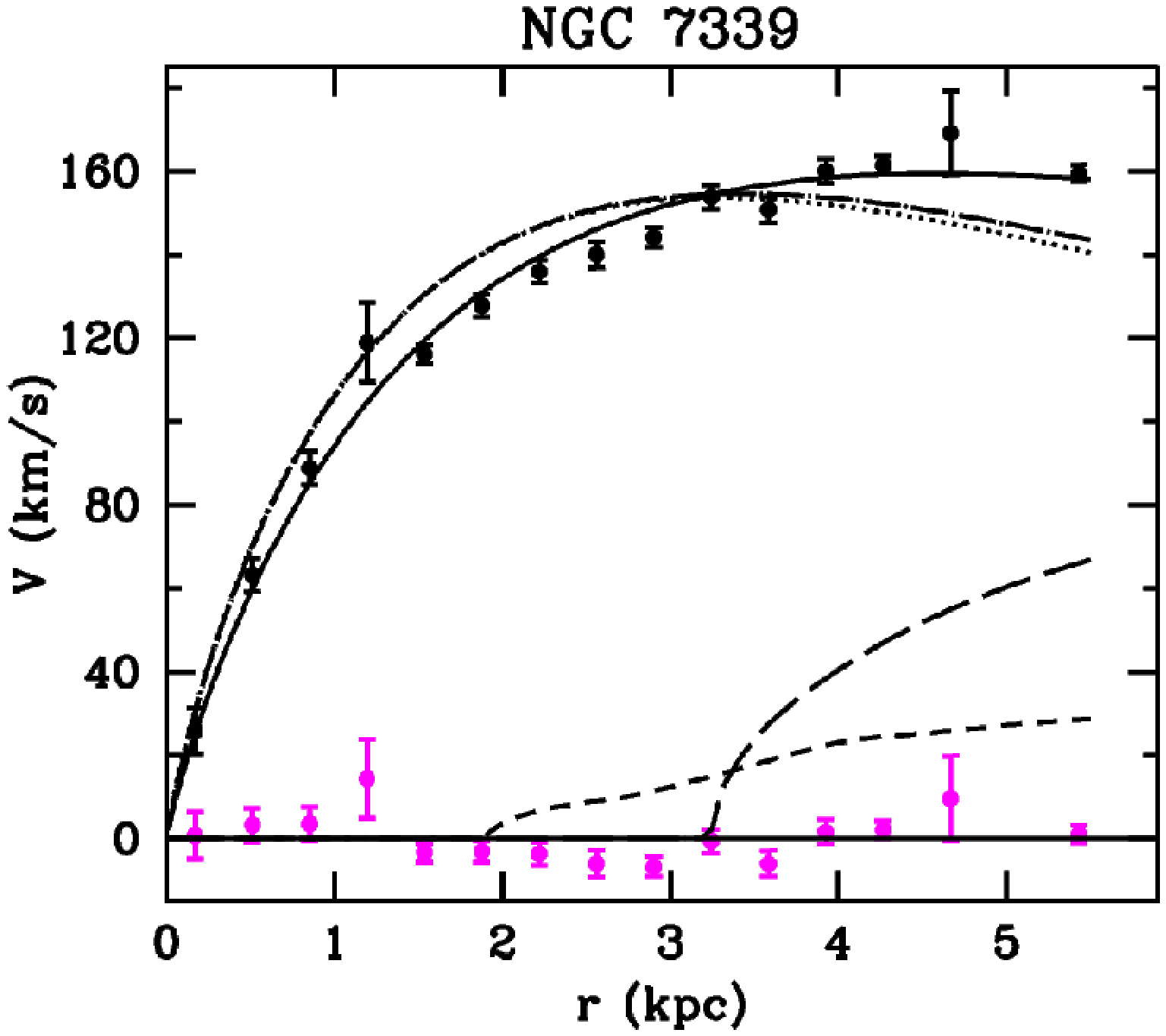}}\goodgap
\subfigure{\includegraphics[width=4.75cm]{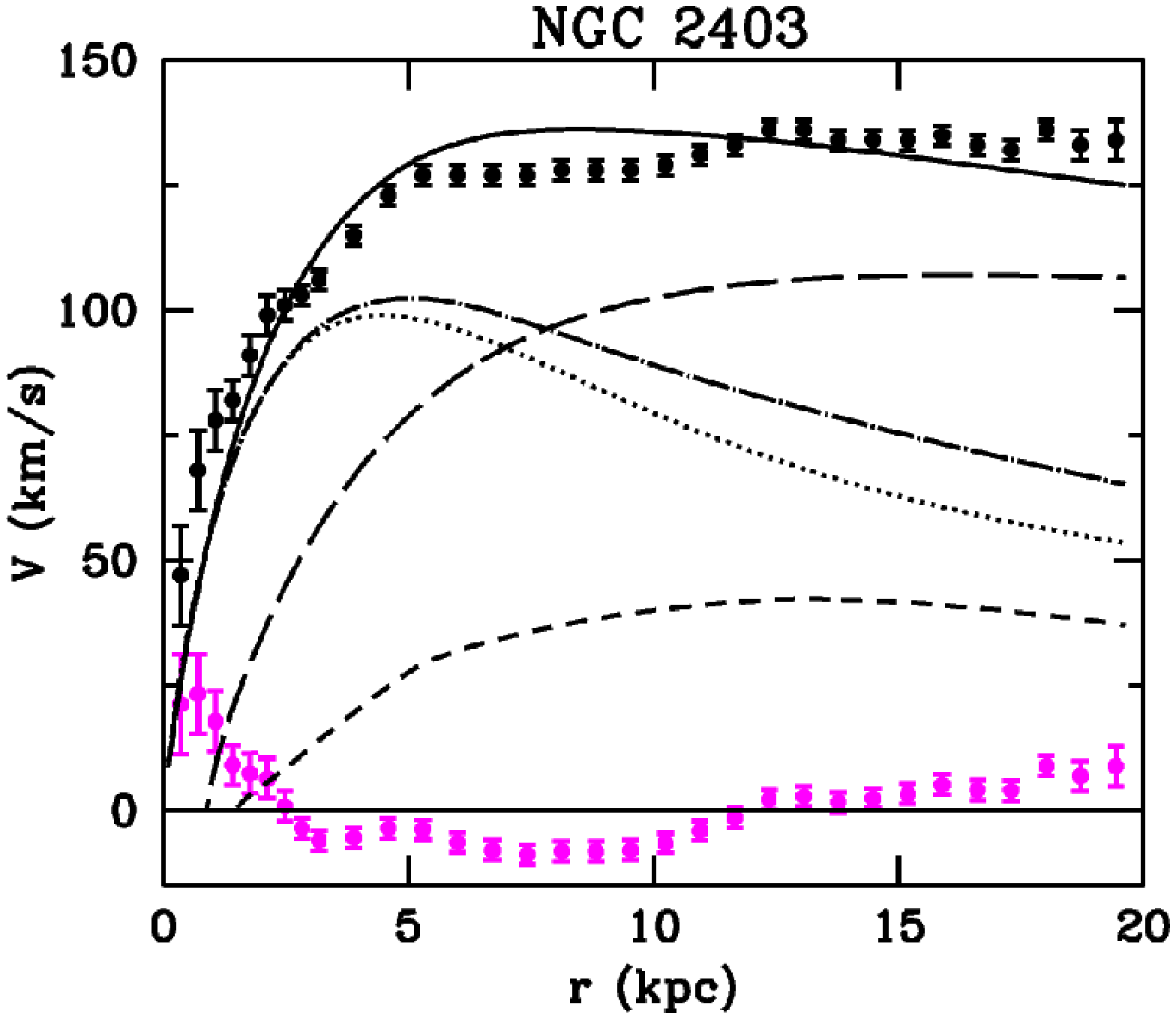}}\goodgap\\
\subfigure{\includegraphics[width=4.75cm]{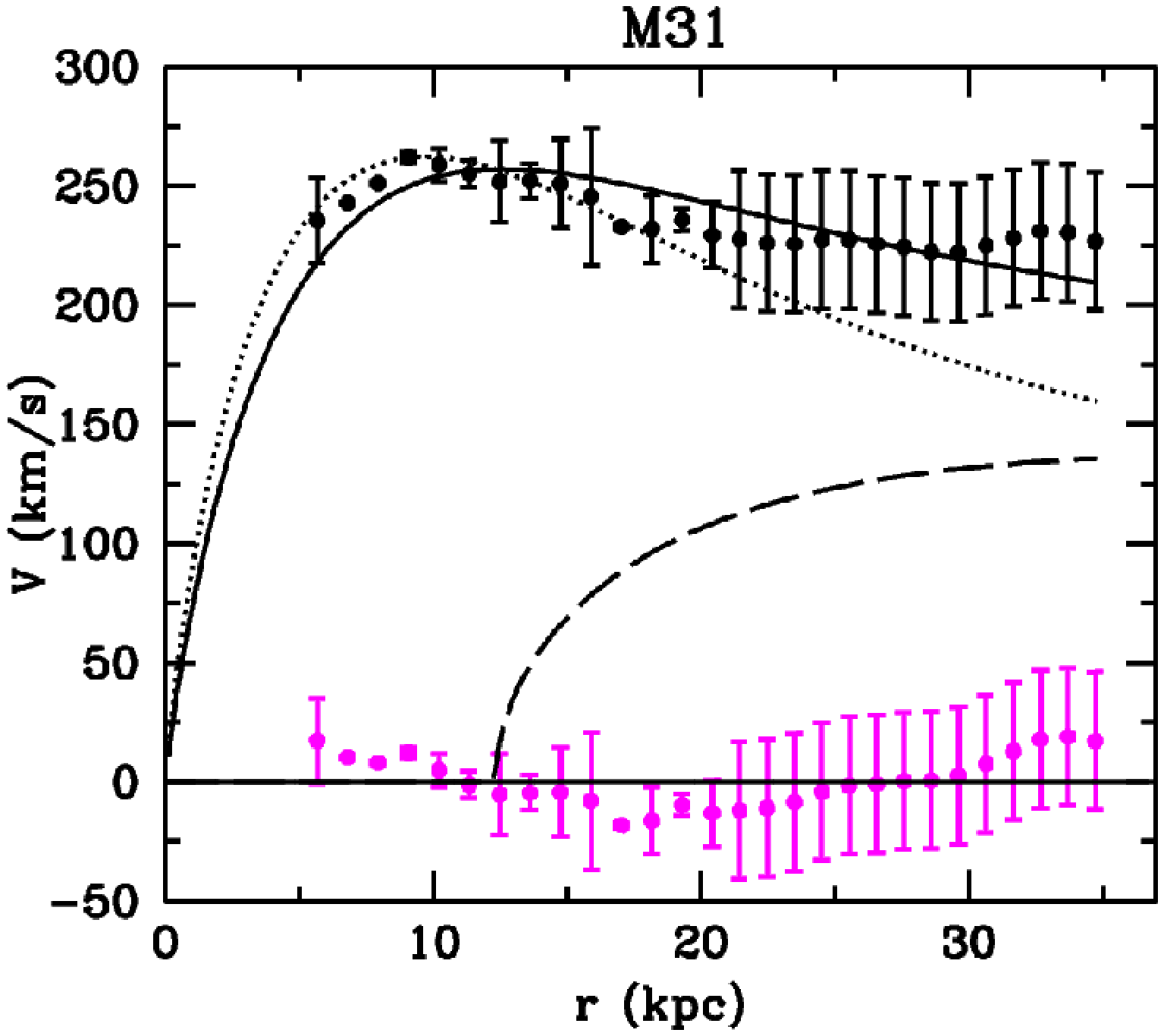}}\goodgap
\subfigure{\includegraphics[width=4.75cm]{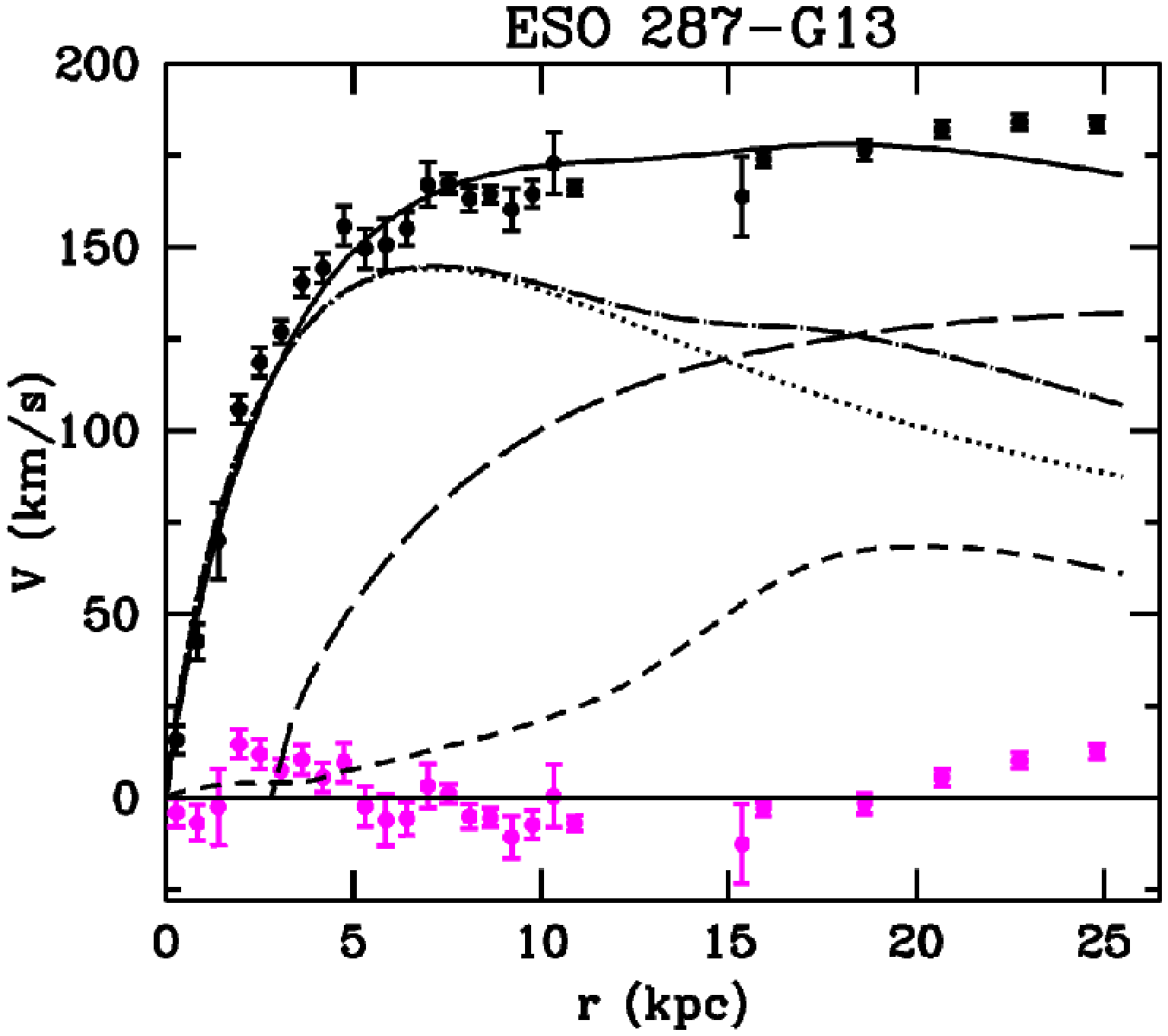}}\goodgap
\subfigure{\includegraphics[width=4.75cm]{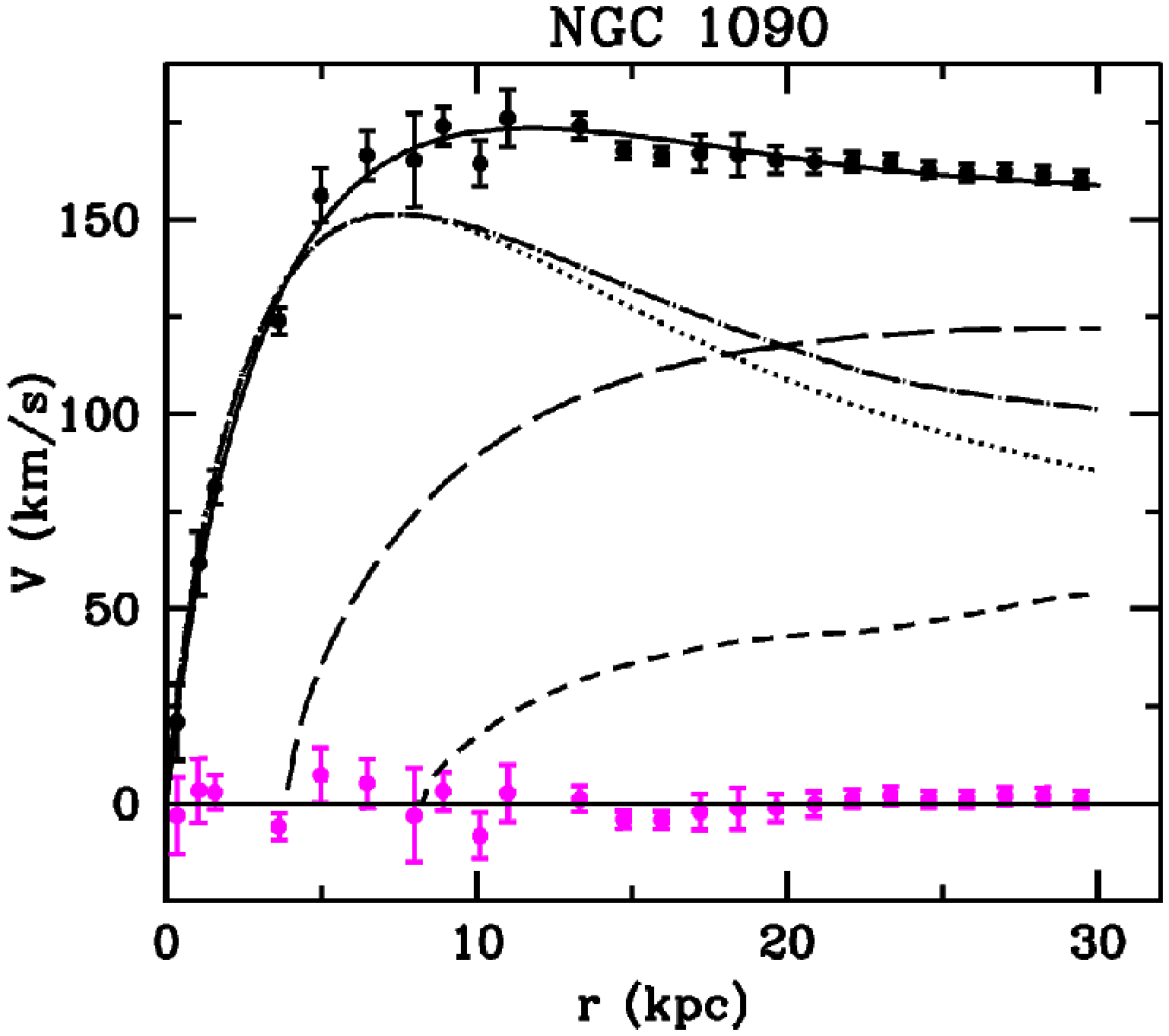}}\goodgap\\
\subfigure{\includegraphics[width=4.75cm]{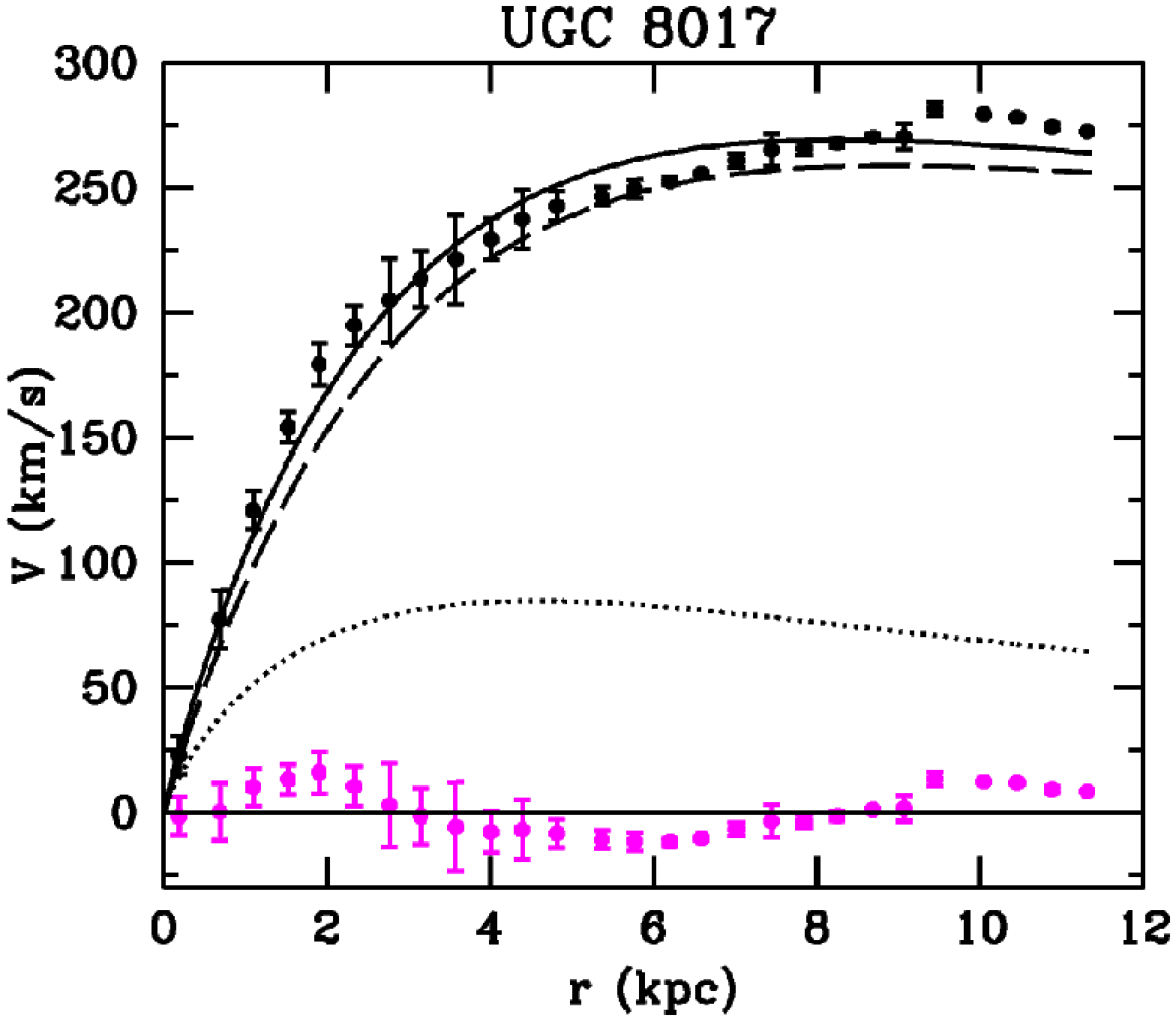}}\goodgap
\subfigure{\includegraphics[width=4.75cm]{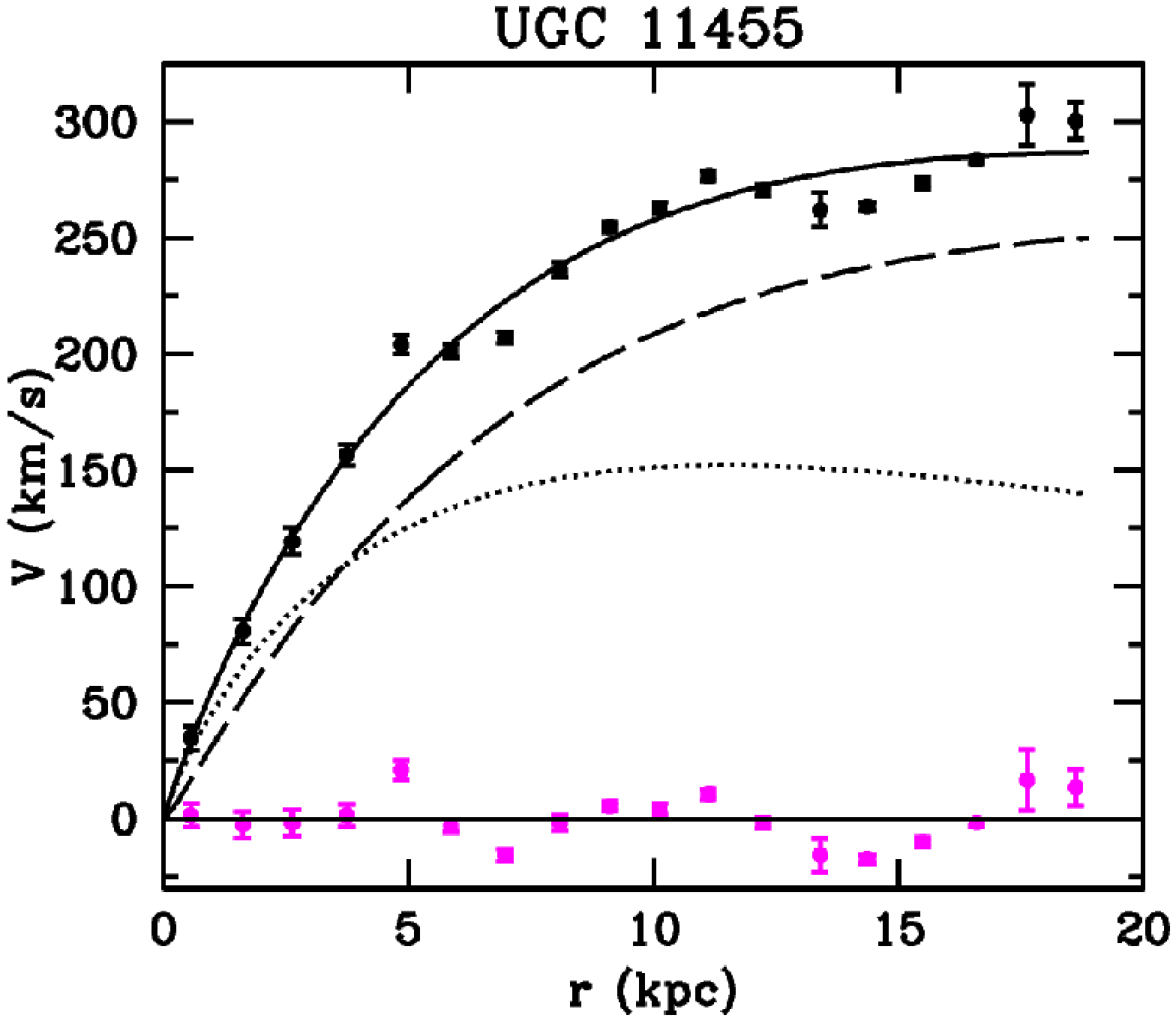}}\goodgap
\subfigure{\includegraphics[width=4.75cm]{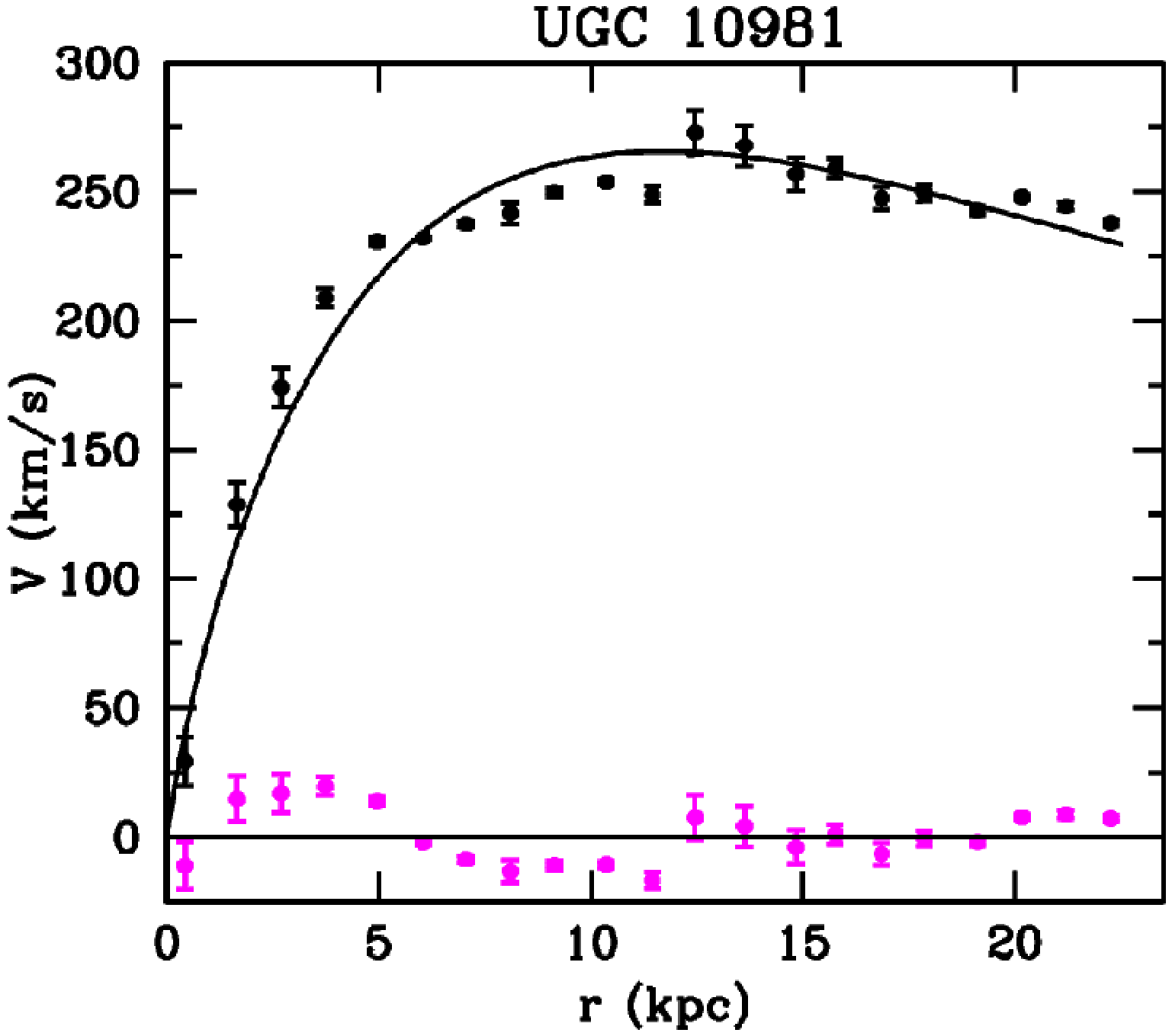}}\goodgap\\
\caption{\emph{Sample A}: The solid line represents the best-fit total circular velocity $V_{CCT}$.
The dashed and dotted lines are the Newtonian contributions from the gas and the stars, while the dot-dashed represents their sum.
The long-dashed line is the non-Newtonian contribution of the gas and the stars to the model.
Below the  RCs, we plot the residuals ($V_{obs}-V_{CCT}$).
See Table 1 for details.}
\label{fig:bf}
\end{figure*}

\begin{figure*}
\centering
\subfigure{\includegraphics[width=4.75cm]{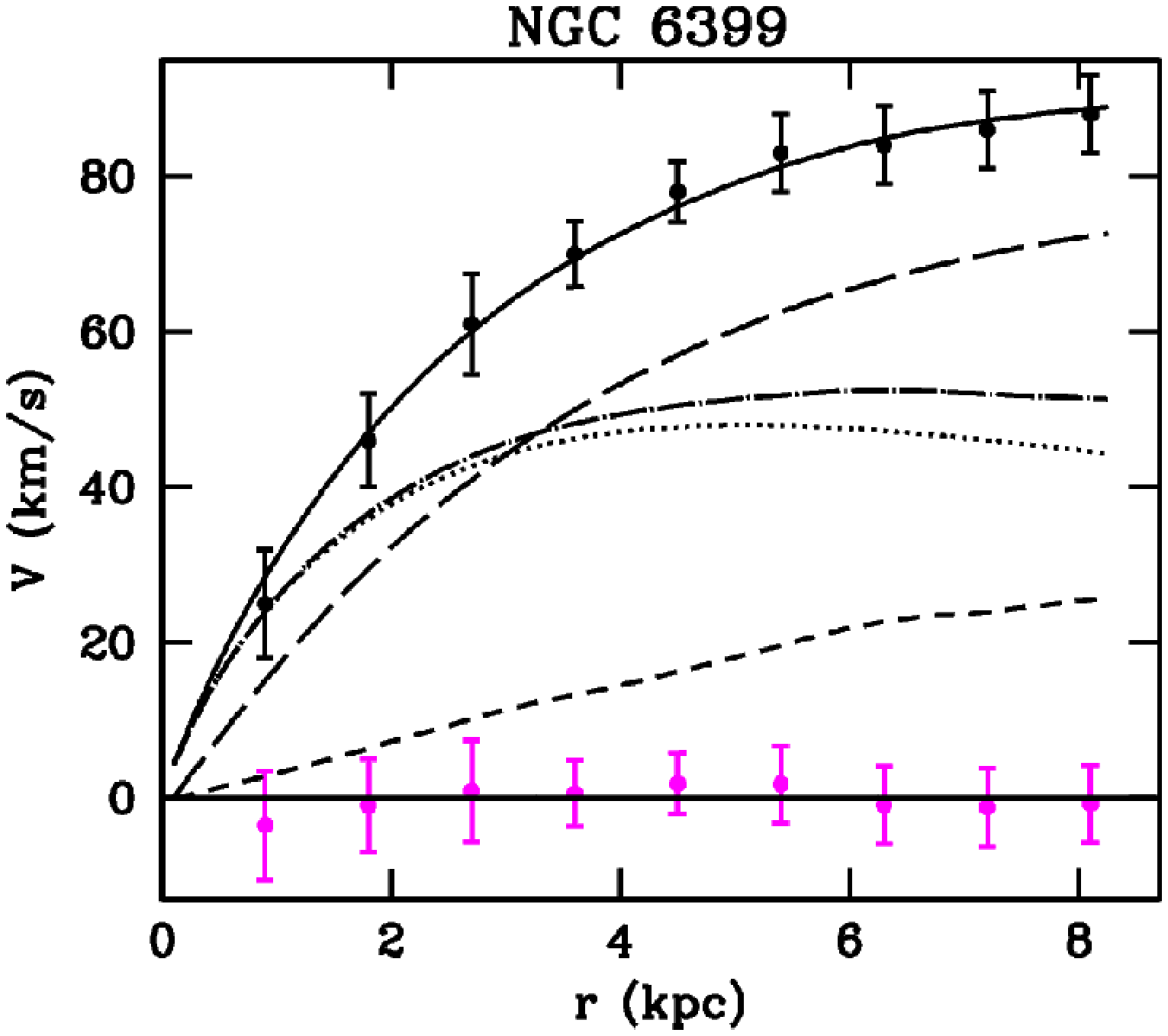}}\goodgap
\subfigure{\includegraphics[width=4.75cm]{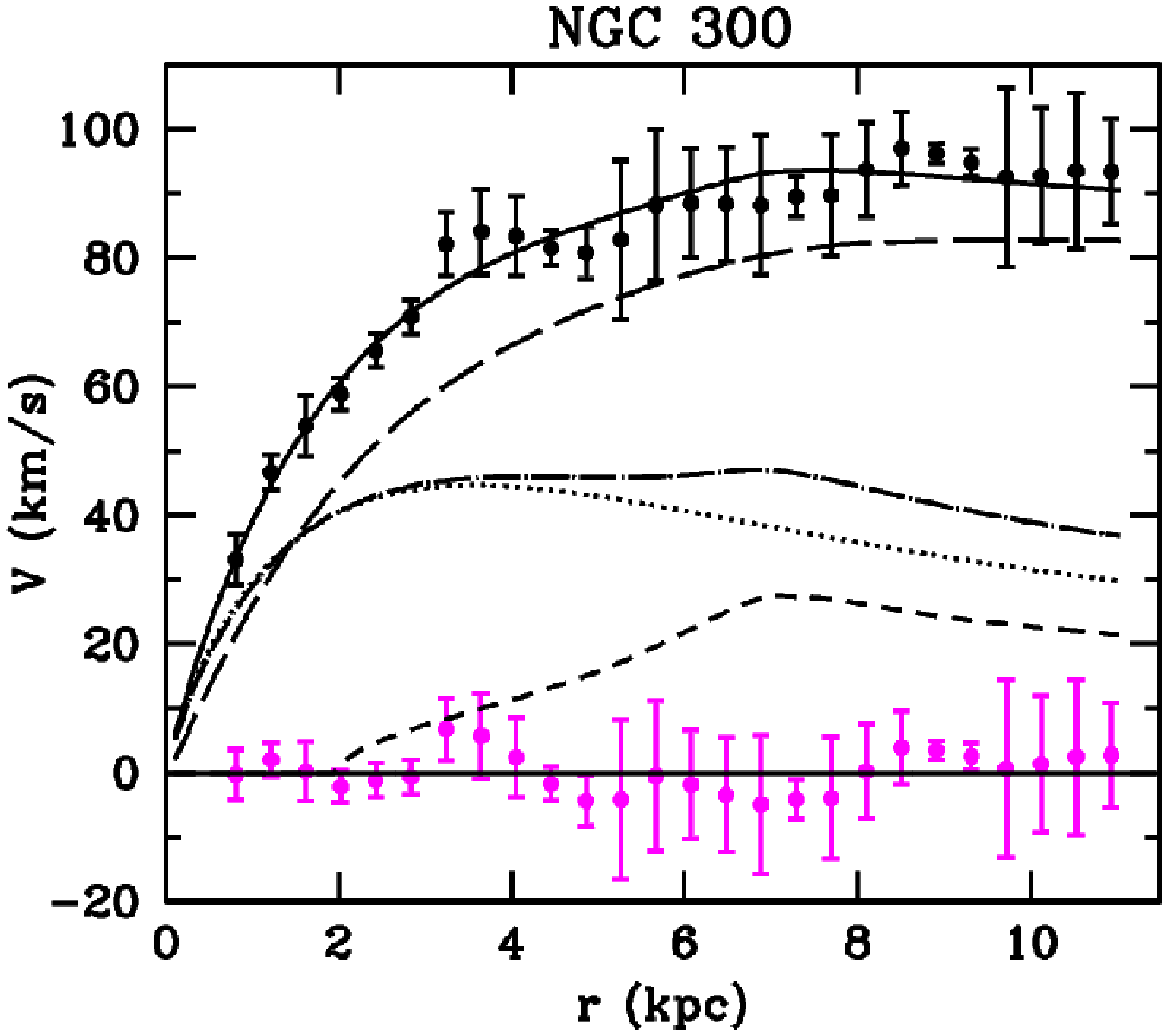}}\goodgap
\subfigure{\includegraphics[width=4.75cm]{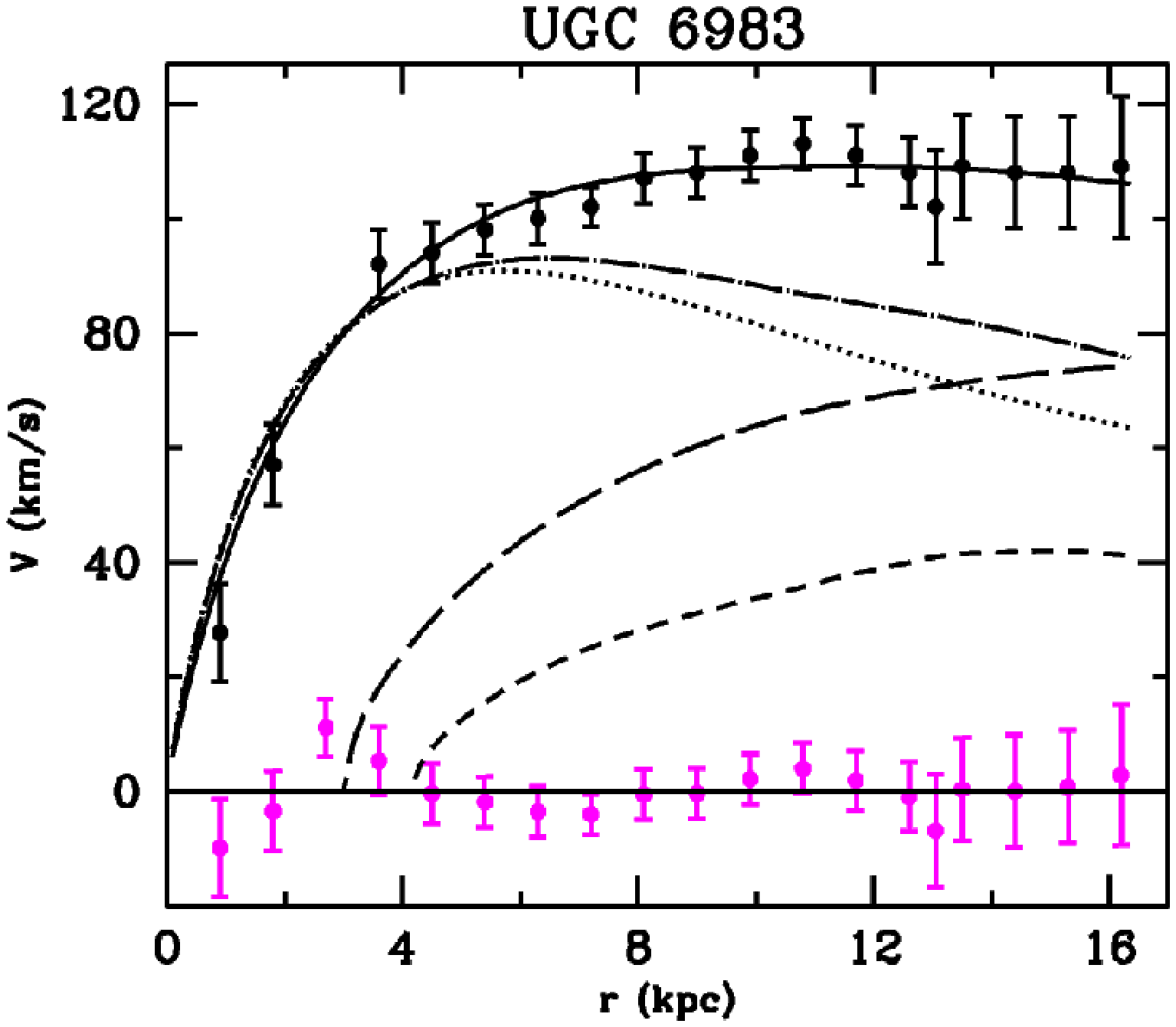}}\goodgap\\
\subfigure{\includegraphics[width=4.75cm]{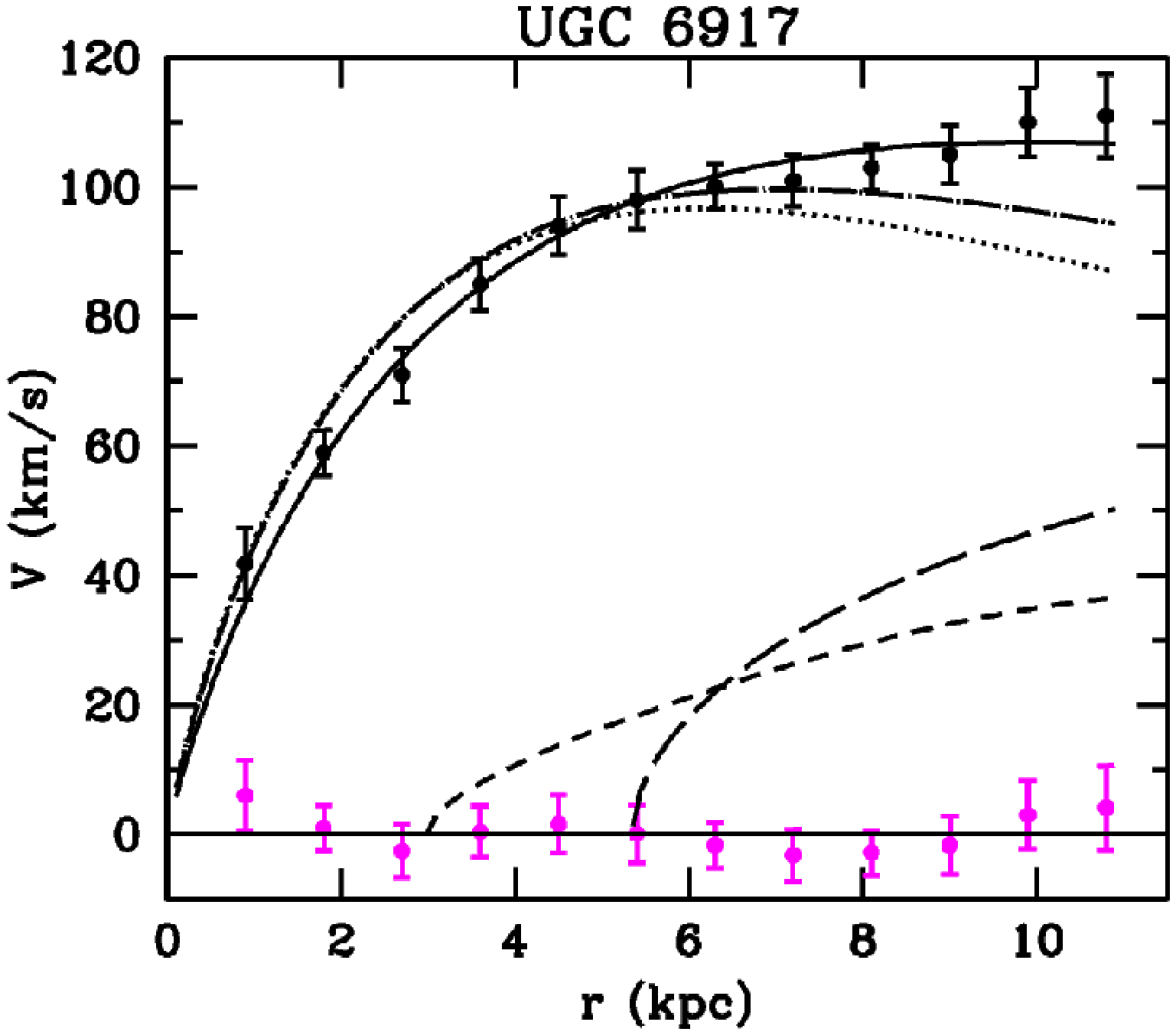}}\goodgap
\subfigure{\includegraphics[width=4.75cm]{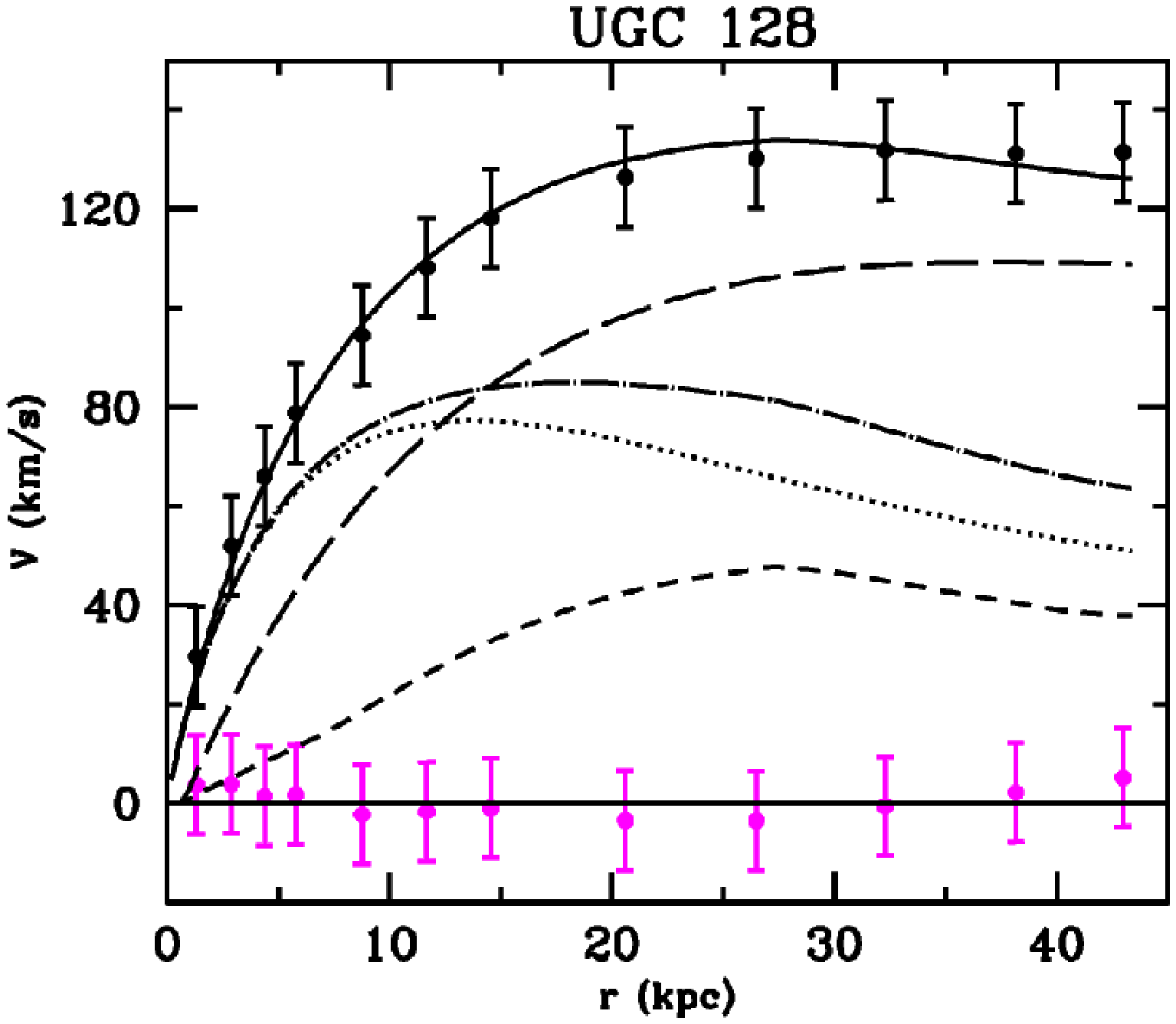}}\goodgap
\subfigure{\includegraphics[width=4.75cm]{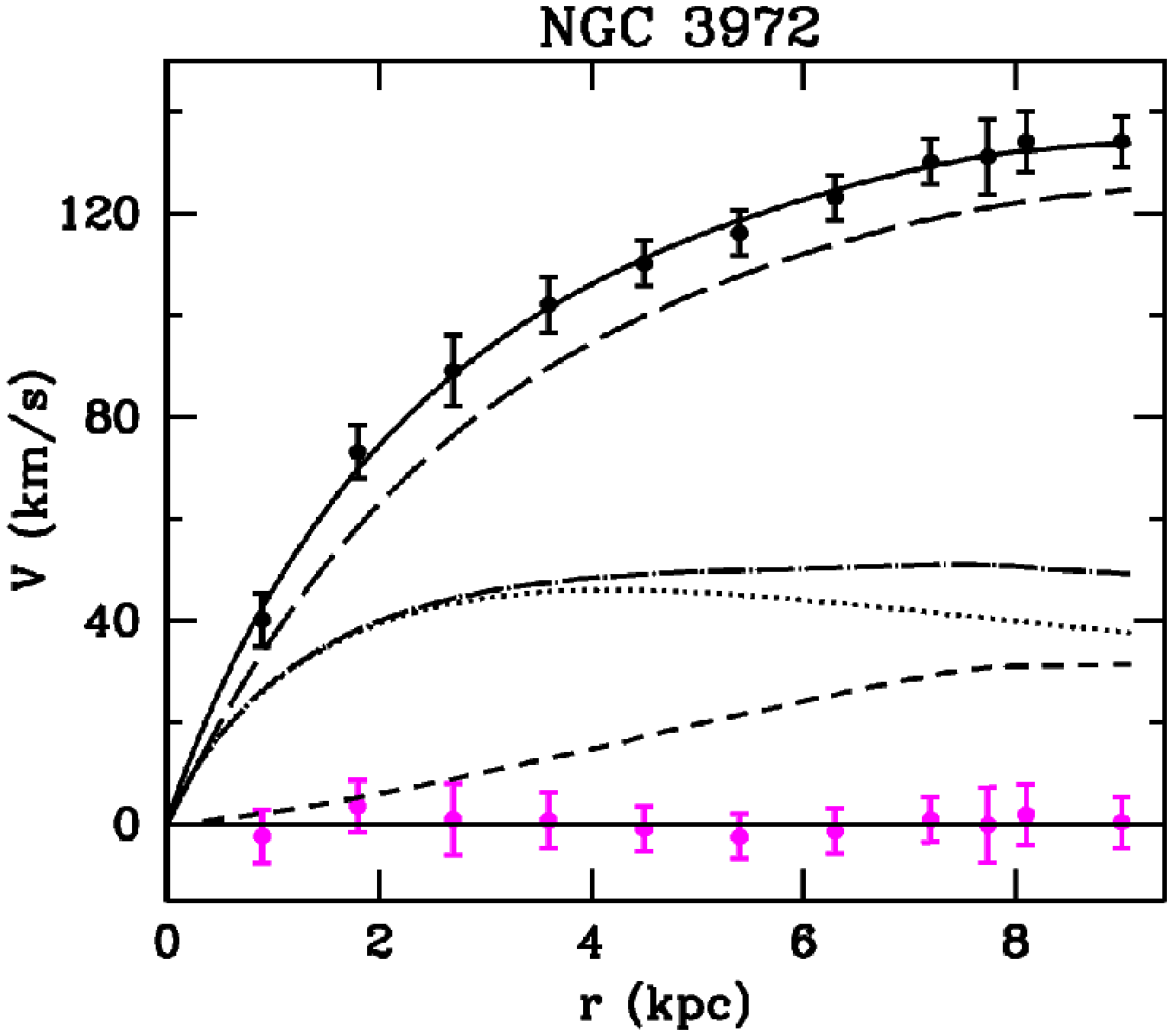}}\goodgap\\
\subfigure{\includegraphics[width=4.75cm]{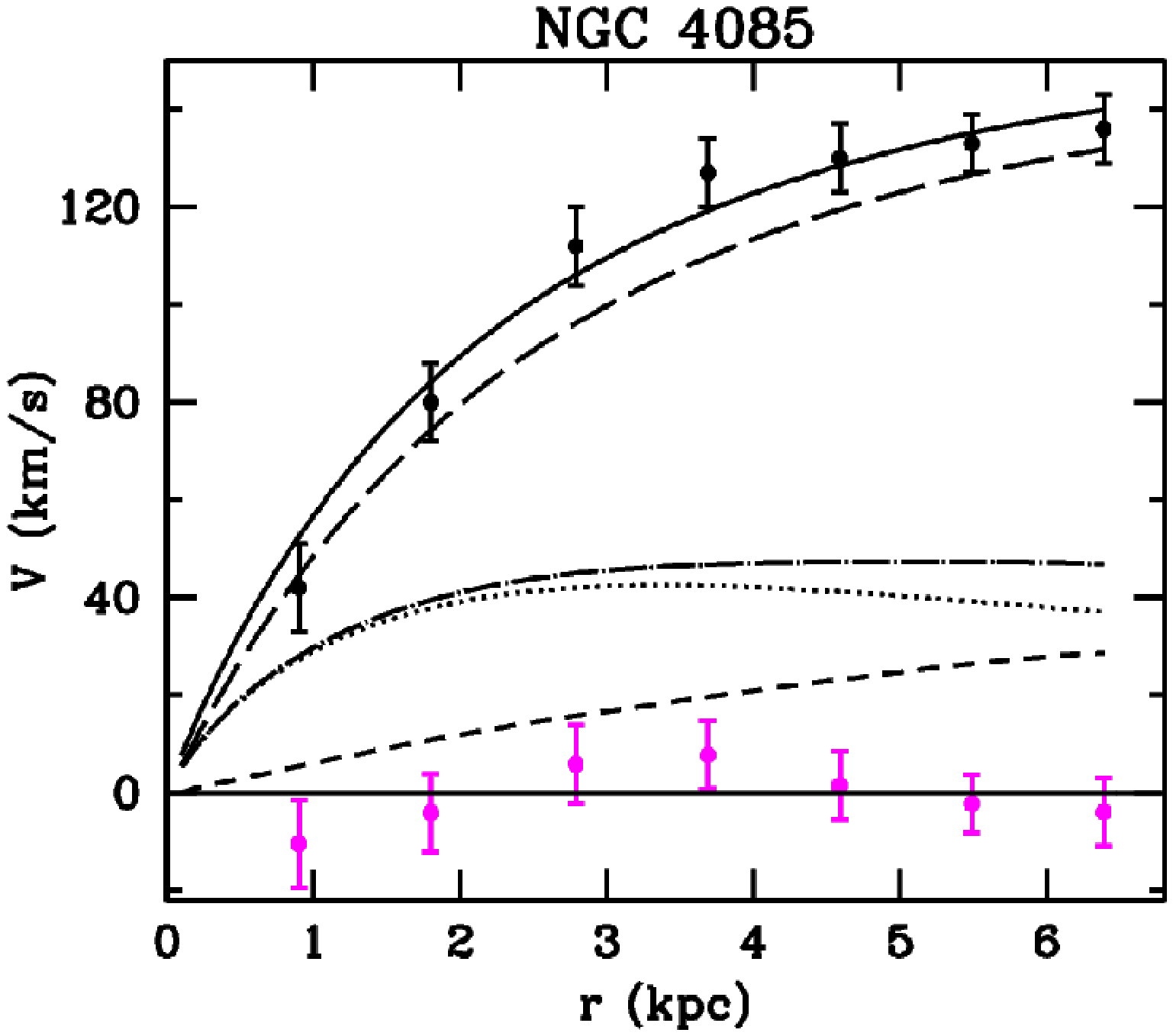}}\goodgap
\subfigure{\includegraphics[width=4.75cm]{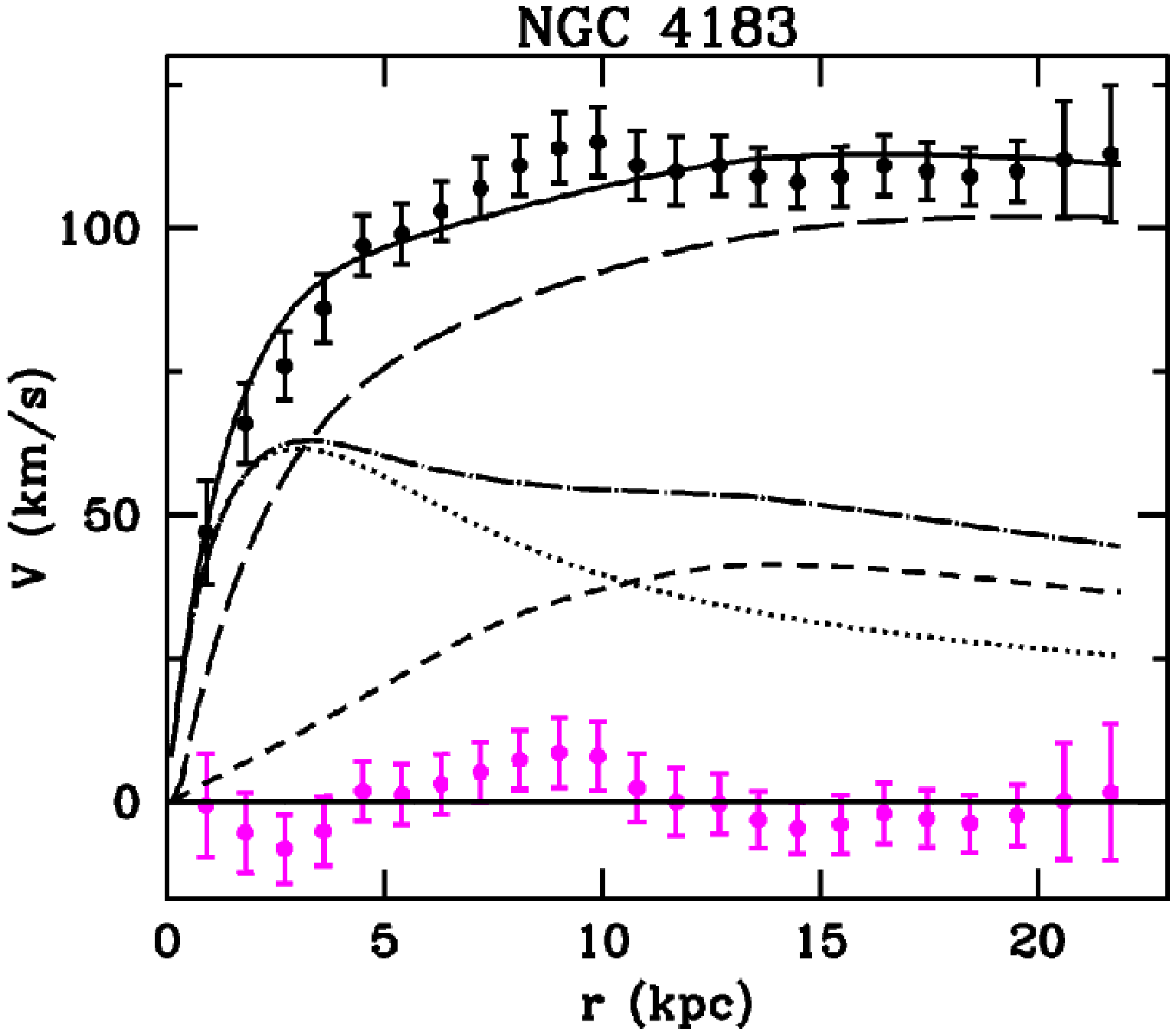}}\goodgap
\subfigure{\includegraphics[width=4.75cm]{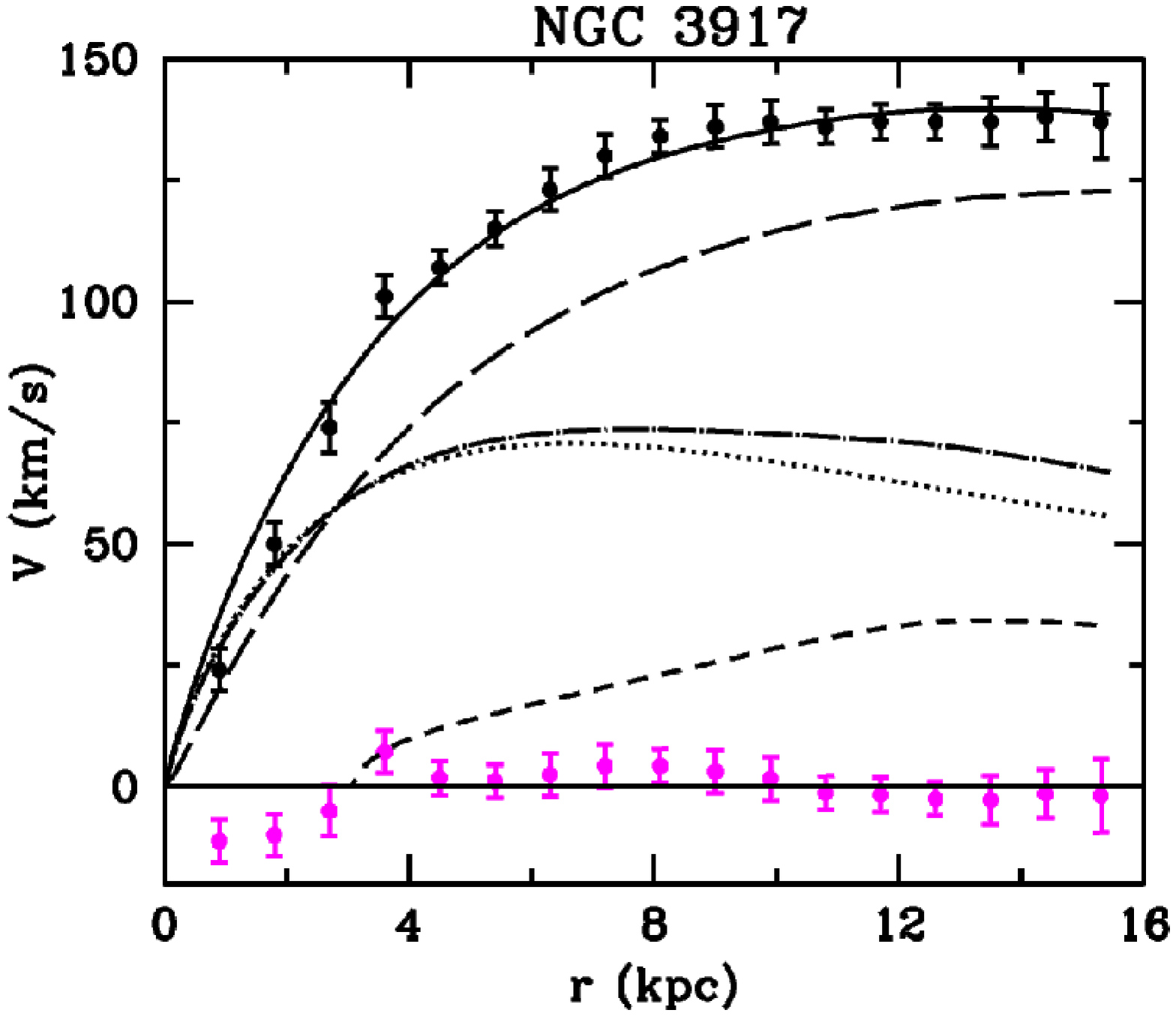}}\goodgap\\
\subfigure{\includegraphics[width=4.75cm]{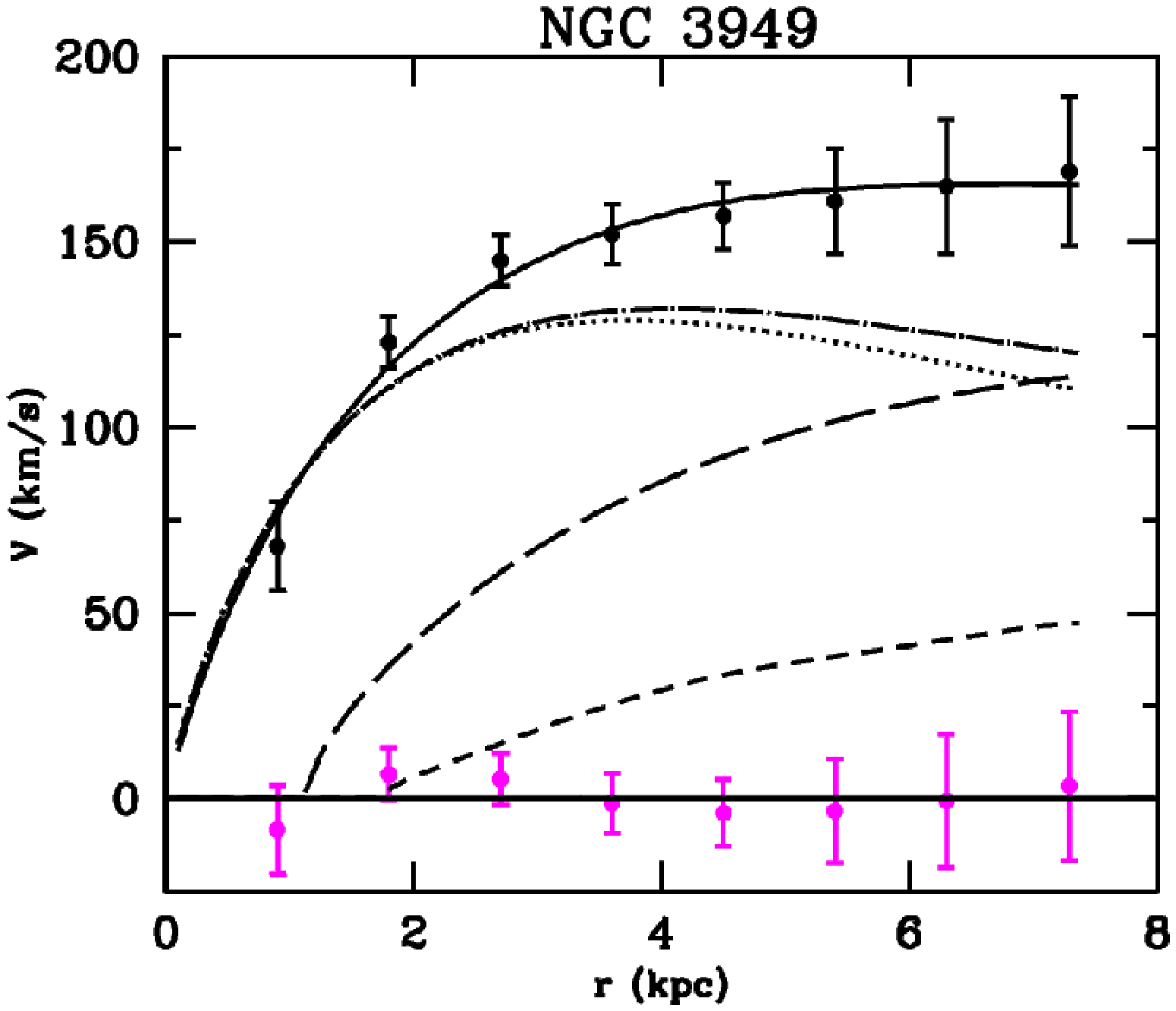}}\goodgap
\subfigure{\includegraphics[width=4.75cm]{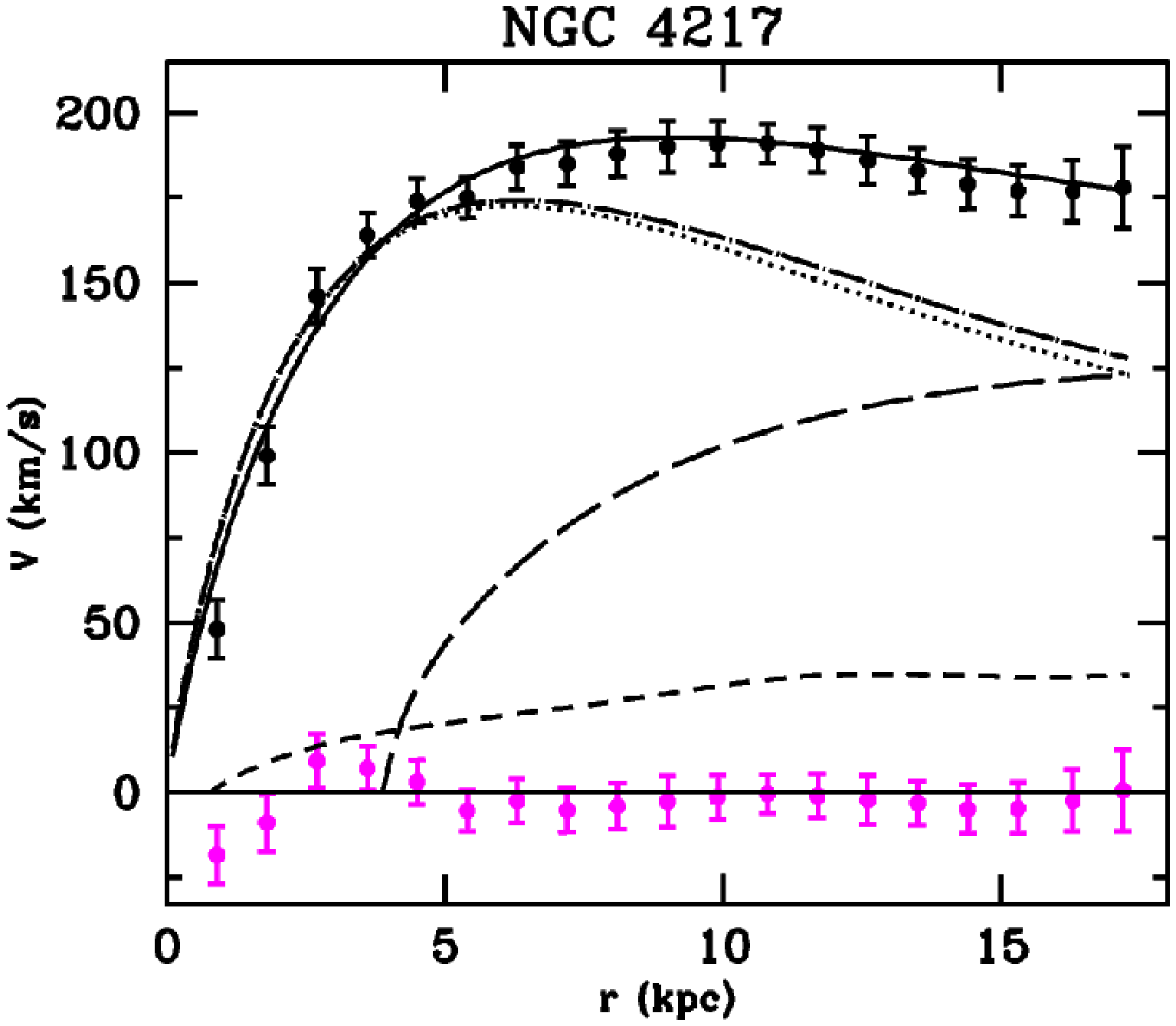}}\goodgap
\subfigure{\includegraphics[width=4.75cm]{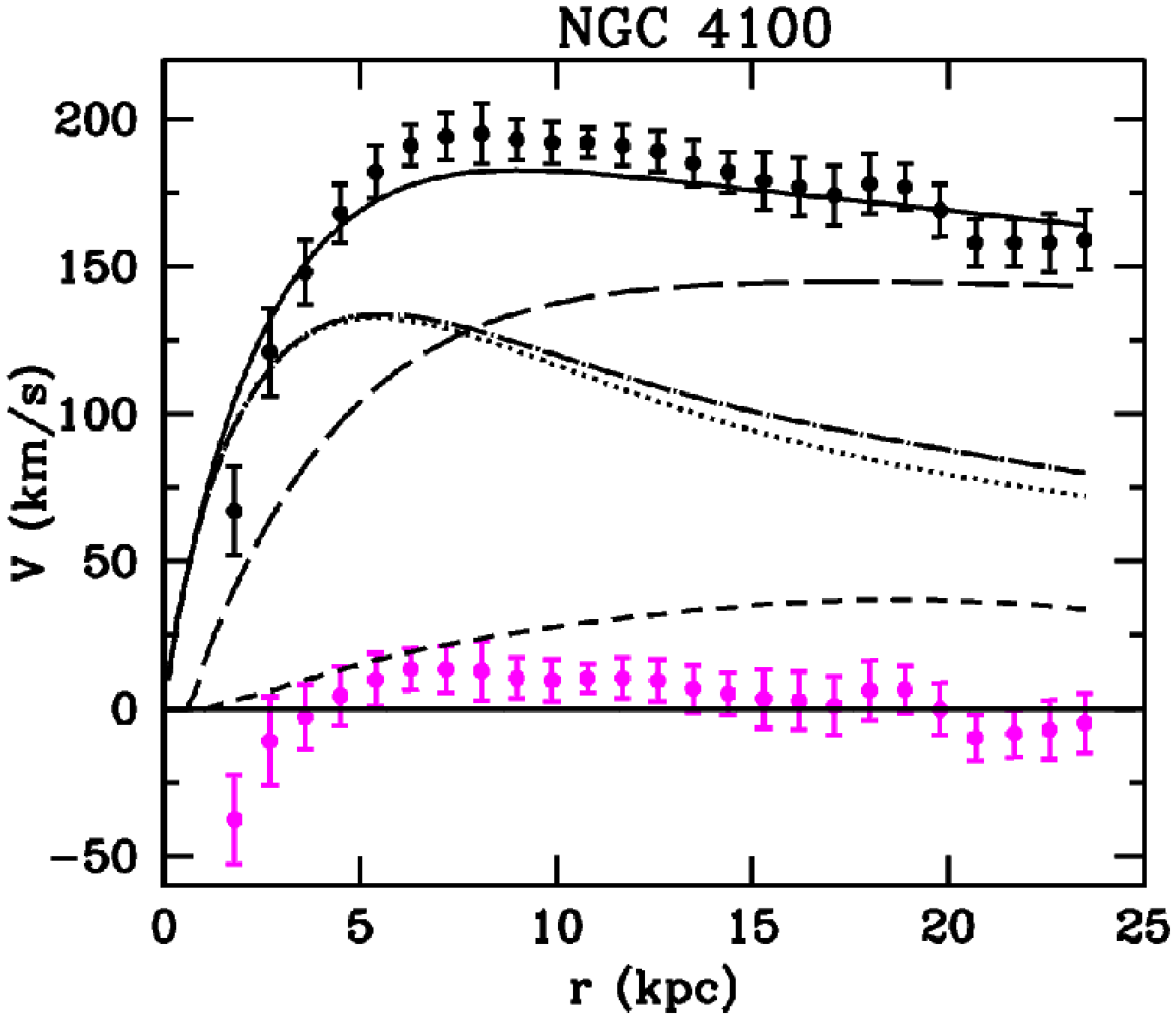}}\goodgap\\
\subfigure{\includegraphics[width=4.75cm]{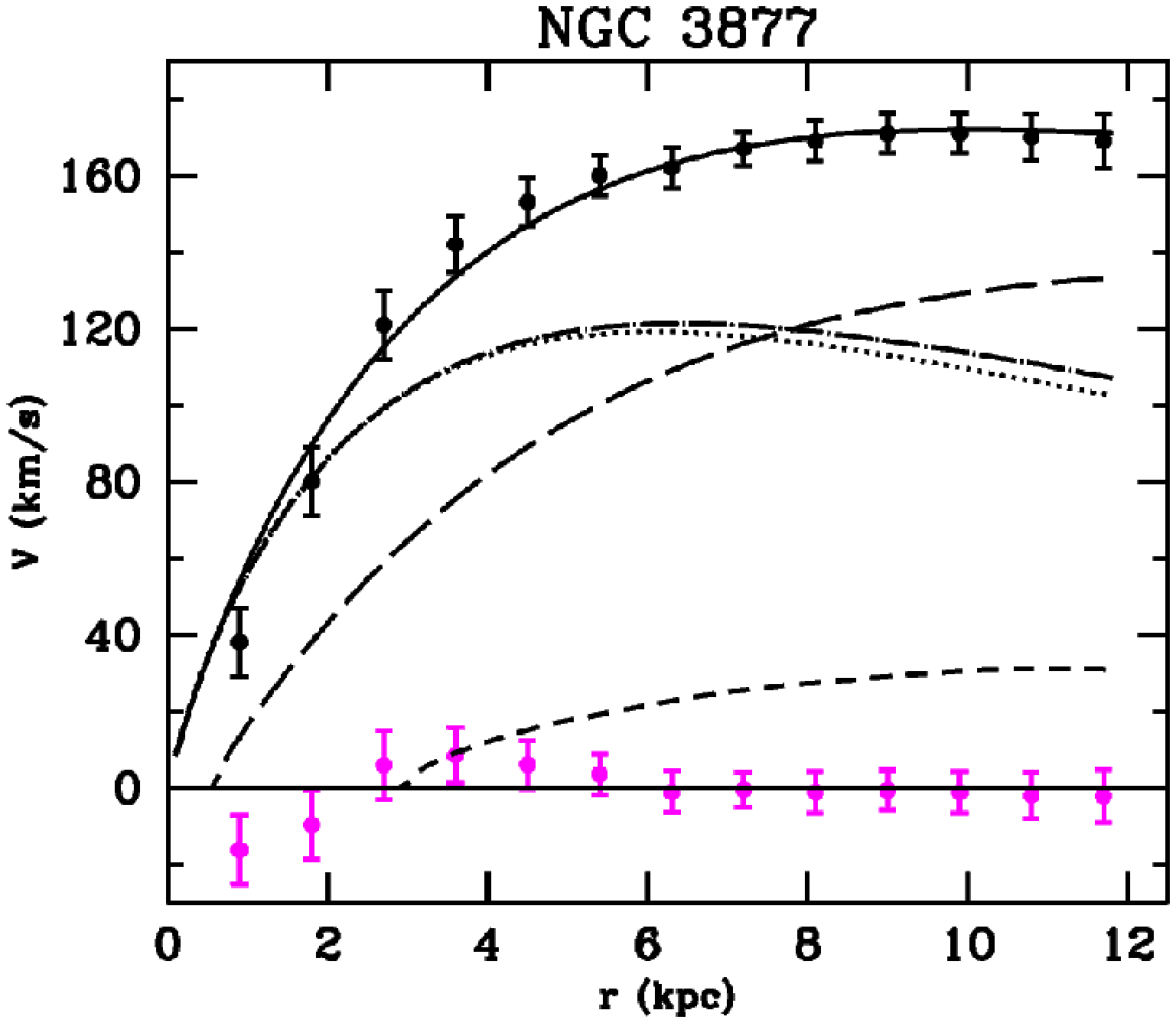}}\goodgap
\subfigure{\includegraphics[width=4.75cm]{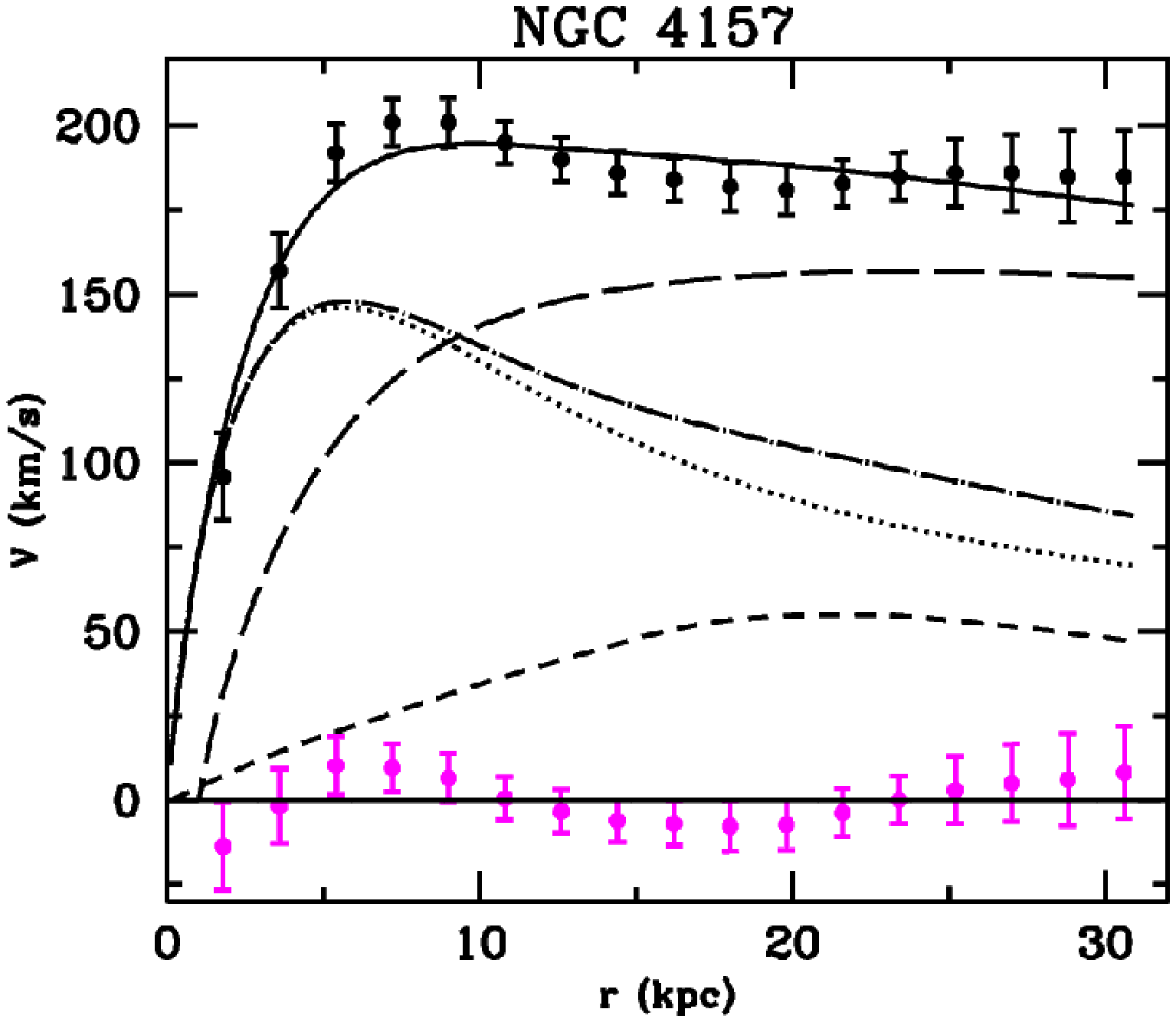}}\goodgap
\subfigure{\includegraphics[width=4.75cm]{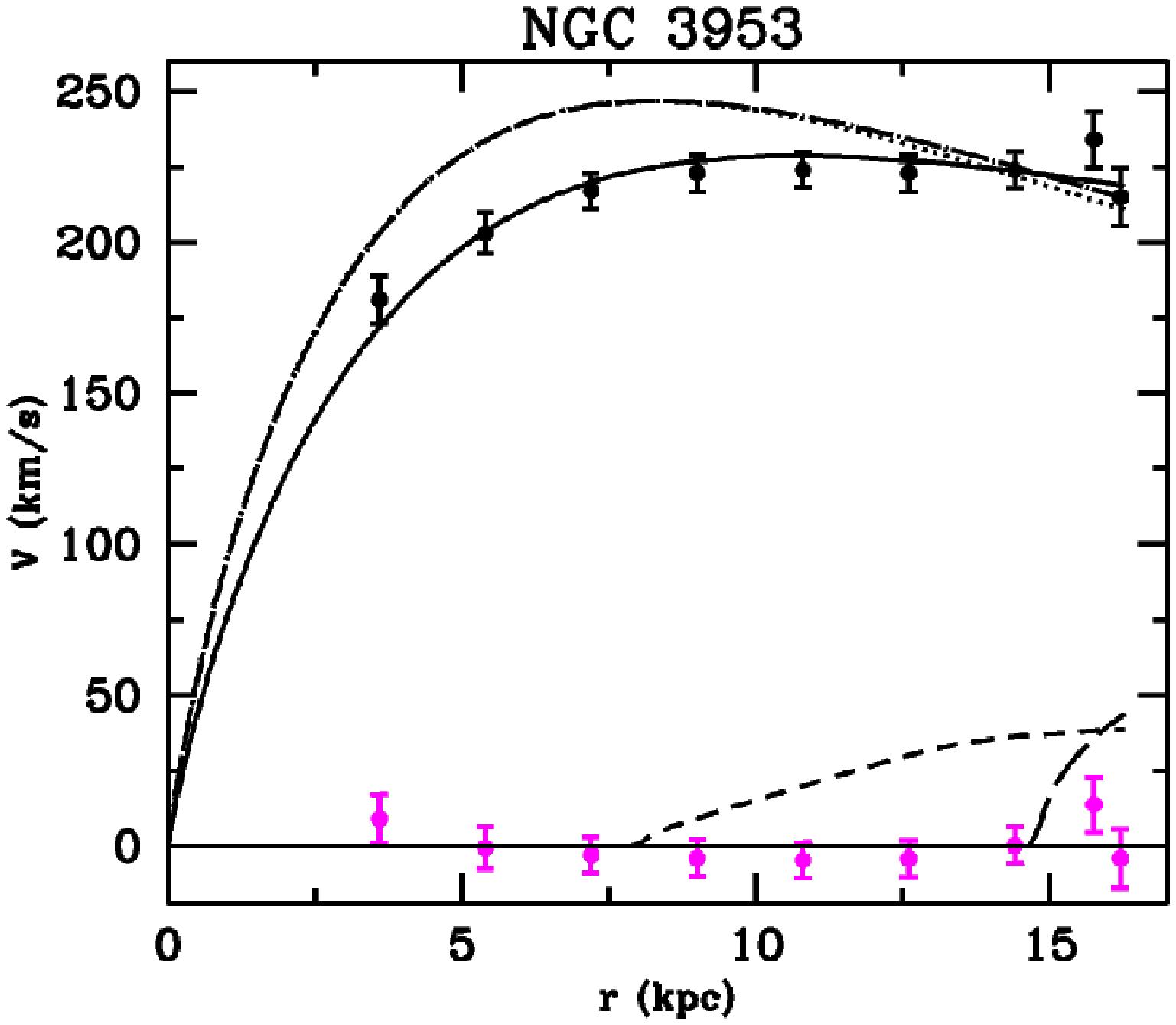}}\goodgap\\
\caption{\emph{Sample B}: Best-fit curves superimposed to the data from
  selected objects from Sanders \& McGaugh 2002. See Figure 1 for details.}
\end{figure*}

\begin{table*}
\caption{Properties and parameters of the mass model of the analyzed Samples ($\beta=0.7$).
From left to right, the columns read: name of the galaxy, Hubble type as reported in the NED database, adopted distance in $Mpc$, B-band luminosity in $10^{9}L_{B \odot}$, disk scale length in $kpc$, gas mass in $10^{9} M_{\odot}$ until last measured point, gas fraction in $\%$, disk mass in $10^{9} M_{\odot}$, scale length CCT parameter in $kpc$, mass-to-light ratio in $\Upsilon_{\odot}^{B}$, and $\chi^{2}_{red}$.
The galaxies are ordered from top to bottom with increasing luminosity.}
\begin{center}
\begin{tabular}{l|c|c|c|c|c|c|c|c|c|c|c|c}\hline\hline
\emph{Galaxy}&\emph{Type}&\emph{D}&\emph{$L_{B}$}&\emph{$R_{D}$}&\emph{$M_{gas}$}&{$f_{gas}$}&\emph{$M_D$}&\emph{$r_c$}&\emph{$\Upsilon_{\star}^{B}$}&\emph{$\chi_{red}^2$}\\ \hline\hline
\multicolumn{11}{c}{\emph{Sample A}}\\\hline\hline
DDO 47&IB&4&0.1&0.5&2.2&96$\pm$1&0.01&0.005&0.1&0.5\\\hline
IC 2574&SABm&3&0.8&1.78&0.52&79$\pm$12&0.14&0.017$\pm$0.003&0.2&0.8\\\hline
NGC 5585&SABc&6.2&1.5&1.26&1.45&58$\pm$3&1&0.038$\pm$0.004&0.7&1.4\\\hline
NGC 55&SBm&1.6&4&1.6&1.3&84$\pm$7&0.24&0.024$\pm$0.004&0.06&0.14\\\hline
ESO 116-G12&SBcd&15.3&4.6&1.7&21&50&2.1&0.05$\pm$0.01&0.5&1.2\\\hline
NGC 6503&Sc&6&5&1.74&2.3&18$\pm$0.7&10.6&0.21$\pm$0.014&2.1&18\\\hline
M 33&Sc&0.84&5.7&1.4&3.7&53$\pm$2&3.3&0.075$\pm$0.004&0.58&25\\\hline
NGC 7339&SABb&17.8&7.3&1.5&6.2&2.8$\pm$0.2&22&0.41$\pm$0.07&3&2.3\\\hline
NGC 2403&Sc&3.25&8&2.08&4.46&27$\pm$0.9&12.1&0.21$\pm$0.015&1.5&19\\\hline
M 31&Sb&0.78&20&4.5&-&-&180$\pm$70&1.53$\pm$0.19&9&3.4\\\hline
ESO 287-G13&Sbc&35.6&30&3.3&14&25$\pm$1&41&0.48$\pm$0.05&1.4&3.2\\\hline
NGC 1090&Sbc&36.4&38&3.4&100&18$\pm$1&47&0.59$\pm$0.04&1.2&0.9\\\hline
UGC 8017&Sab&102.7&40&2.1&-&-&9.1$\pm$0.3&0.01$\pm$0.01&0.23&5.2\\\hline
UGC 11455&Sc&75.4&45&5.3&-&-&74$\pm$3&0.14$\pm$0.01&1.6&5\\\hline
UGC 10981&Sbc&155&120&5.4&-&-&460$\pm$200&$\sim 10^{11}$&3.8&4.9\\\hline\hline
\multicolumn{11}{c}{\emph{Sample B}}\\\hline\hline
UGC 6399&Sm&18.6&1.6&2.4&1&23$\pm$3&3.3&0.1$\pm$0.03&2&0.1\\\hline
NGC 300&Scd&1.9&2.3&1.7&1.3&39$\pm$4&2&0.052$\pm$0.010&0.87&0.43\\\hline
UGC 6983&SBcd&18.6&4.2&2.7&4.1&24$\pm$2&13&0.46$\pm$0.1&3.1&0.88\\\hline
UGC 6917&SBd&18.6&4.4&2.9&2.6&14$\pm$1&16&0.71$\pm$0.17&3.6&0.47\\\hline
UGC 128&Sd&60&5.2&6.4&10.7&32$\pm$5&23&0.39$\pm$0.11&4.4&0.1\\\hline
NGC 3972&Sbc&18.6&6.7&2&1.5&39$\pm$3&2.5&0.025$\pm$0.004&0.37&0.1\\\hline
NGC 4085&Sc&18.6&6.9&1.6&1.3&44$\pm$4&1.7&0.014$\pm$0.003&0.25&1\\\hline
NGC 4183&Scd&18.6&9.5&1.4&4.9&60$\pm$6&3.2&0.09$\pm$0.023&0.3&0.33\\\hline
NGC 3917&Scd&18.6&11&3.1&2.6&22$\pm$1.5&9.2$\pm$0.9&0.098$\pm$0.014&0.8&1\\\hline
NGC 3949&Sbc&18.6&19&1.7&4.1&19$\pm$2.2&17&0.22$\pm$0.06&0.9&0.25\\\hline
NGC 4217&Sb&18.6&21&2.9&3.3&6.1$\pm$0.7&52&0.55$\pm$0.15&2.5&0.38\\\hline
NGC 4100&Sbc&18.6&25&2.5&4.4&13$\pm$1.5&28&0.20$\pm$0.03&1.1&1.52\\\hline
NGC 3877&Sc&18.6&27&2.8&1.9&7.3$\pm$0.8&24&0.2$\pm$0.04&0.9&0.75\\\hline
NGC 4157&Sb&18.6&30&2.6&12&26$\pm$2.6&33&0.25$\pm$0.04&1.1&0.53\\\hline
NGC 3953&SBbc&18.6&41&3.8&4&2.8$\pm$0.18&140&1.9$\pm$0.5&3.4&0.78\\\hline\hline
\end{tabular}
\end{center}
\end{table*}

\end{document}